\documentclass[english,floatfix,superscriptaddress,twocolumn,showpacs,aps,prb,reprint, longbibliography]{revtex4-2}

\usepackage{amsmath}
\usepackage{graphicx}
\usepackage{bm}
\usepackage{dsfont}
\usepackage{xfrac}
\usepackage{siunitx}
\usepackage{rotating}
\usepackage[resetlabels]{multibib}
\newcites{sm}{References}
\usepackage{xcolor} 
\newcommand{\vect}[1]{\boldsymbol{\mathbf{#1}}}
\newcommand{\dg}{^{\dagger }}

\newcommand{\bR}{{\bf{R}}}

\renewcommand{\Re}{\operatorname{Re}}
\renewcommand{\Im}{\operatorname{Im}}
\usepackage{hyperref}
\hypersetup{colorlinks=true, citecolor=blue, urlcolor=blue, linkcolor=blue, breaklinks=true}

\begin{document}

\title{Tuning magnetic interactions with nonequilibrium optical phonon populations: a theoretical study}


\author{Milan Kornja\v ca}
\author{Rebecca Flint}
\affiliation{Department of Physics and Astronomy, Iowa State University, 12 Physics Hall, Ames, Iowa 50011, USA}
\date{\today}
\begin{abstract}
We theoretically explore how light-driven optical phonons can be used to drive magnetic exchange interactions into interesting physical regimes by developing a general theory of spin-phonon pumping in magnetic insulators with non-equilibrium optical phonon distributions, focusing on the diabatic regime where phonon frequencies are much larger than the magnetic interactions. We present several applications of spin-phonon pumping two-dimensional nearest-neighbor Heisenberg, XYZ and Kitaev models to examine what kind of further neighbor interactions and chiral fields can be generated, and how anisotropic couplings can be enhanced, showing that experimentally accessible non-equilibrium phonon distributions can generically drive significant frustration and realize a variety of spin liquid regimes. This effect is described for both direct and superexchange mechanisms, and we derive simple geometric rules for which phonon modes are ``spin-phonon'' active and for which magnetic interactions. Spin-phonon pumping provides an intriguing possibility for preferentially pumping specific magnetic interaction terms. In addition to generating further neighbor interactions, such pumping can lead to increased magnetic anisotropy for initially weakly anisotropic models, and selectively pumping the Kitaev-Heisenberg model can suppress undesirable Heisenberg terms while enhancing Kitaev interactions.

\end{abstract}
\maketitle
\section{Introduction}

The interplay between magnetic and lattice degrees of freedom
plays a vital role in the physics of strongly correlated materials. Spin-phonon coupling has been extensively explored both as a way to
probe ordered phases in frustrated magnets \cite{Fernandes2019, Kreisel2011} and quantum spin liquids \cite{Zhou2011, Serbyn2013, Ye2020,  Metavitsiadis2021, Ferrari2021}, and as a tool for tuning the equilibrium magnetic properties of materials \cite{Kuboki1987, Hase1993, Khomskii1996, Uhrig1998, Weisse1999, Zhang2008}. Recent experimental advances in the light control of materials have opened up the possibility to tune correlated states and magnetism in a non-equilibrium setting by coupling light to different microscopic degrees of freedom. One fruitful avenue has been coupling light to electronic degrees of freedom, which in turn tune the magnetic interactions \cite{Mentink2015, Claassen2017, Oka2019, Marion2019, Quito2021, Quito2021b}.  However, this approach often requires large fluences that can induce significant heating \cite{Alessio2014, Claassen2017}. An alternative route is to pump resonant optically active phonons by infrared (IR) light, which can drive a wide range of correlated phenomena, including metal-insulator transitions \cite{Rini2007} and exotic superconductivity \cite{Mankowsky2014, Mitrano2016, Knap2016}. These non-equilibrium phonon distributions may be also able to effectively tune magnetism via spin-phonon coupling, where changes have already been seen in seminal experiments \cite{Wall2009, Nova2017, Maehrlein2018, Disa2020, Afanasiev2021, Padmanabhan2022, Disa2023}.

The theoretical treatment of these effects, however, has largely been limited to effective classical descriptions, first principle calculations, and some group-theoretical arguments \cite{Afanasiev2021, Padmanabhan2022, Disa2023, Lin2024}, although the quantum effects of phonon pumping have recently been considered in spin-polariton  \cite{Pantazopoulos2024} and magnon-phonon settings  \cite{Yarmohammadi2024}. However, a general, easy to interpret quantum treatment is still lacking.  In this paper, we capture the first-order quantum effects of optical phonon pumping directly on the magnetic exchange interactions in a simple theoretical treatment. We focus on spin-phonon models on two-dimensional (2D) lattices, and describe how quantum effects lead to generic, intuitive rules for the different magnetic couplings that can be induced by phonon pumping. Depending on the lattice geometry, exchange mechanism, and phonon mode choices, we show that experimentally accessible optical phonon pumping can enhance frustration or anisotropy to drive systems into effective magnetic Hamiltonians supporting exotic ground states. 

\begin{figure}[!htb]
\vspace{-.1cm}
\includegraphics[width=0.7\columnwidth]{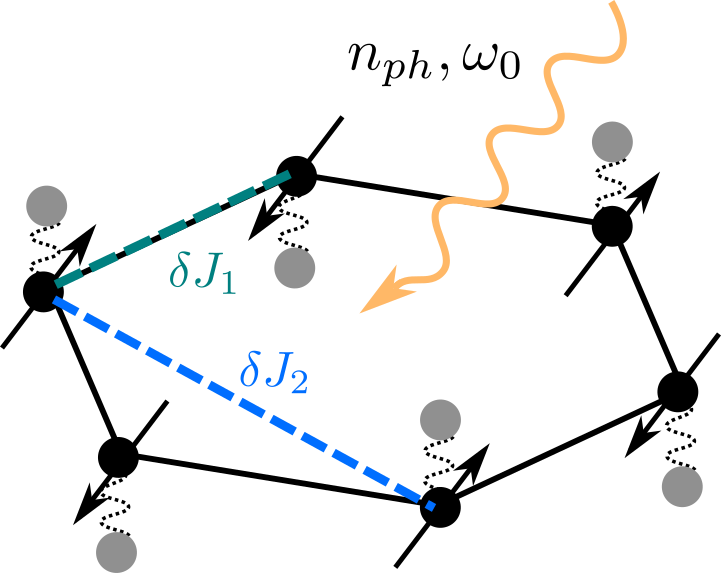}
\caption{
Optical phonons are pumped by IR light, inducing a non-equilibrium $q=0$ phonon distribution, $n_{ph}(\omega_0)$ that dynamically affects the interatomic distances (gray atoms) and induces phonon mediated changes to the magnetic interactions. The leading order quantum effects of this spin-phonon coupling both change the initial couplings, $\delta J_1$ and generate further neighbor couplings, $\delta J_2$.  We also find that the magnetic exchange anisotropy is generically enhanced, and circularly polarized phonon modes can additionally generate chiral fields (not shown). Strong pumping without significant heating is possible for frequencies below the Mott gap, $\omega_0 \ll U$.}\label{fig:Fig1_intro}
\vspace{-0.cm}
\end{figure}

Equilibrium spin-phonon coupling has been explored extensively in one dimension as a paradigmatic model for spin-Peierls transitions \cite{Kuboki1987, Uhrig1998, Weisse1999, Zhang2008}. Here, we generalize this approach to higher dimensions and more general spin interactions using an approximate unitary transformation to integrate out the phonons.  The leading spin-phonon effects significantly modify the interactions in the resulting effective spin model \cite{Weisse1999}. Prior work employed similar approaches to the electron-phonon interaction and phonon pumping, which tunes electronic interactions to induce superconductivity \cite{Kennes2017}.  Our approach specifically considers the quasi-stationary regime where relevant pumping timescales are longer than the phonon distribution transients, allowing the driven phonon modes to be described by a non-equilibrium distribution, $n_{ph}(\omega_0)$ \cite{Yarmohammadi2021, Allafi2024}.  As long as the magnetic energy scales are below the phonon frequency, the spins will adiabatically evolve in the background of this non-equilibrium phonon occupancy \cite{Weisse1999, Hager2007}.  This induced phonon population acts as dynamical lattice distortion that affects the exchange couplings through their strong dependence on the atomic locations, and can both change existing and generate new magnetic interactions, as shown schematically in Fig. \ref{fig:Fig1_intro}.

This work presents a general framework for calculating leading quantum non-equilibrium phonon pumping effects on magnetic exchange interactions. Representative results are derived for simple 2D lattices, where we demonstrate a clear advantage compared to electronically mediated pumping, as sizable magnetic frustration can be induced with currently accessible pumped phonon occupancies. This advantage is supplemented by diminished expected heating effects on the electronic subsystem, as the light drive directly heats only the phonon bath with frequencies well below the electronic energy scales. We develop simple geometrical rules to predict the driven phonon modes most effective for magnetic exchange tuning, and apply them to more complex lattices and anisotropic spin models, showcasing the tunability achievable by phonon drive. In particular, specific spin interactions can be preferentially enhanced over others by carefully choosing the phonon mode, amplitude, and polarization. The complexity of real materials provides a zoo of phonon modes to drive, all with potentially different effects. This potential to tune frustration \cite{Balents2010, Savary2016} and anisotropy \cite{Jackeli2009, Winter2017, Hermans2018} by driving specific phonon modes opens new avenues for realizing exotic states of matter like quantum spin liquids in real materials. 

\subsection{Structure of the paper}
We develop a spin-phonon model for magnetic interaction tuning via phonon pumping in Sec. \ref{sec:Model}, and comment on experimental feasibility in Sec. \ref{sec:SPpump}, with estimates for experimentally achievable coupling constants and pumped phonon occupancies. In Sec. \ref{sec:Results}, we present several examples of what changes in magnetic interactions are possible, examining pumped 2D lattices with both direct and superexchange mechanisms, as well as Heisenberg, XYZ and Kitaev exchange interactions.
Finally, we provide conclusions and outlook in Sec. \ref{sec:conclusions}.

\section{Spin-phonon models and effective spin interactions\label{sec:Model}}
Magnetic exchange interactions in insulators decay rapidly as a function of bond distance due to the exponential dependence of the underlying hopping integrals. Thus, lattice vibrations can lead to strong dynamical fluctuations of the exchange couplings, captured by a spin-phonon Hamiltonian. In this section, we perturbatively treat the leading contribution of the dynamic fluctuations \cite{Kennes2017}, resulting in an effective spin model with modified magnetic exchange couplings.  We do this by performing a unitary transformation on the spin-phonon Hamiltonians and effectively integrating out the phonon degrees of freedom to find the modified effective spin Hamiltonian.

\subsection{The linear spin-phonon Hamiltonian\label{sec:SPham}}

The Hamiltonian has spin ($H_s$), phonon ($H_p$) and the linear spin-phonon interaction ($H_{sp}$) terms \cite{Pytte1974, Bray1975, Kuboki1987,Uhrig1998}:
\begin{equation}\label{eq:H0}
H=H_s+H_{p}+H_{sp}.
\end{equation}

For simplicity, we take nearest-neighbor Heisenberg interactions,  although the approach is readily generalized to different spin Hamiltonians (see Sec. \ref{sec:anisotropic}):
\begin{equation}\label{eq:Hs}
H_s=\sum_{\langle ij\rangle}J_1^{(0)} \vect{S}_i \cdot \vect{S}_j.
\end{equation}

We allow the possibility of multiple phonon modes ($\eta$) with dispersions $\omega_{q\eta}$ and polarizations $\hat \epsilon_{q\eta}$: 
\begin{equation}\label{eq:Hp}
H_p=\sum_{\vect{q}\eta}\omega_{\vect{q}\eta} b\dg_{\vect{q}\eta}b_{\vect{q}\eta}.
\end{equation}
The leading linear spin-phonon interaction is found by Taylor expanding the dependence of $J(\bR)$ on the lattice displacements between nearest-neighbors, $u_{ij}$ \cite{Pytte1974, Bray1975}. We comment on the effect of higher-order terms in the expansion in Sec. \ref{sec:LFtransf}.
\begin{equation}\label{eq:Hsp}
H_{sp}=\sum_{\langle ij\rangle}J_1^{(0)} \lambda u_{ij} \vect{S}_i \cdot \vect{S}_j,
\end{equation}
where the relevant lattice displacements are,
\begin{equation}\label{eq:uij}
u_{ij}=\sum_{\vect{q}\eta}\frac{ e^{-i\vect{q}\cdot \vect{r}_i}}{\sqrt{2M \omega_{\vect{q}\eta} N }}\left(f_{\vect{q}\eta}^{ij}b_{\vect{q}\eta}+f_{-\vect{q}\eta}^{*ij}b\dg_{-\vect{q}\eta}\right).
\end{equation}
Here $M$ is the ionic mass, $N$ the number of sites, and the $f$ parameters are the projection of 
the phonon polarizations onto the relevant direction, $\hat{r}_0^{ij}$, as given below.  $\hat{r}_0^{ij}$ is often along the bond direction, but can be more complicated depending on the underlying mechanism. $\lambda$ is the dimensionful spin-phonon coupling constant:
\begin{equation}\label{eq:lamder}
\lambda=\frac{1}{J_1^{(0)}}\frac{dJ_1(\vect{r})}{d r_0}.
\end{equation}
The $f$'s take different forms depending on the exchange mechanism. For direct exchange, the relevant direction $\hat r_0^{ij}$ is the one changing the bond length\cite{Pytte1974}:
\begin{equation}\label{eq:r0HC}
    \hat{r}_0^{ij}=-\frac{\vect{r}_i-\vect{r}_j}{\left|\vect{r}_i-\vect{r}_j\right|},
\end{equation}
with the $f$ parameters given by:
\begin{equation}\label{eq:fexchange}
        f_{\vect{q}\eta}^{ij}=\hat{r}_0^{ij}\cdot\left(\hat{\epsilon}_{\vect{q}\eta}^i-\hat{\epsilon}_{\vect{q}\eta}^j e^{-i\vect{q}\cdot \left(\vect{r}_i-\vect{r}_j\right)}\right),
\end{equation}
and superexchange mediated through (possibly several) anions (labeled by $A$), with \cite{Feldkemper2000, Rosch2004, Rocquefelte2012}:
\begin{equation}\label{eq:fsuperex}
        f_{\vect{q}\eta}^{ij}=\sum_{A_{\langle ij\rangle}}\hat{r}_0^{ij}\cdot\hat{\epsilon}_{\vect{q}\eta}^{A_{\langle ij\rangle}}.
\end{equation}
The relevant directions for the superexchange paths are discussed in more detail in \ref{sec:SEpaths}.

\subsection{Lang-Firsov transformation\label{sec:LFtransf}}

These spin-phonon Hamiltonians can be analytically treated using the Lang-Firsov unitary transformation,
\begin{equation}\label{eq:cantransf}
  \tilde{H}=e^{S}H e^{-S},
\end{equation}
which decouples the spin and phonon degrees of freedom to first order in the spin-phonon coupling \cite{Uhrig1998, Weisse1999, Yasuda2015}. This transformation was used to treat spin-phonon effects in equilibrium 1D spin models relevant for CuGeO$_3$ \cite{Hase1993,Khomskii1996,Akiyama2011}. Here, we focus on the phonon distribution dependent terms relevant for pumping.

In the general Lang-Firsov treatment \cite{Uhrig1998}, we use a variational unitary transformation,
\begin{align}\label{eq:LFtransf}
    S[g_{\vect{q}}]=&\!\!\sum_{\vect{q}\eta, \langle ij \rangle}\!\frac{g_{\vect{q} }J_1^{(0)}\lambda}{\omega_{\vect{q}\eta}}u_{ij} \vect{S}_i \cdot \vect{S}_j \\
    =&\!\!\sum_{\vect{q}\eta, \langle ij \rangle}\!\!\!\frac{g_{\vect{q} }J_1^{(0)} \lambda e^{-i\vect{q}\cdot \vect{r}_i}}{\sqrt{2M N\omega_{\vect{q}\eta}^3}} \left(f_{-\vect{q}\eta}^{*ij}b\dg_{-\vect{q}\eta}-f_{\vect{q}\eta}^{ij}b_{\vect{q}\eta}\right)\vect{S}_i \cdot \vect{S}_j,\notag
\end{align}
where $g_{\vect{q}}$ is a variational parameter chosen such that the first order coupling between spins and phonons, $\langle H_{sp} + \left[S[g_{\vect{q}}], H_s+H_p\right] \rangle_H$ is minimized when averaged over the ground state of the full Hamiltonian. We focus mainly on the \emph{diabatic} regime, defined by $J_{1}^0<\omega_{q\eta}$, where the phonon dynamics are \emph{fast} compared to the spins. Due to the difference in energy scales, the phonons can be considered free, while the spins feel the average effect of phonon dynamics. Therefore, the phonon average can be taken independently, making $\langle \left[S[g_{\vect{q}}], H_s \right] \rangle=0$, as $\langle b_{q\eta} \rangle_{H_p} =\langle b\dg_{q\eta} \rangle_{H_p}=0$. The optimal $g_{\vect{q}}=1$ then leads to complete cancellation to first order in $\lambda$ with,
\begin{align}\label{eq:cancelation}
    \left[S[1],H_p\right]=-H_{sp}.
\end{align}

In other regimes, the optimal $g_{\vect{q}}$ may be different, most notably in the adiabatic regime where the optimal $g_{\vect{q}}$ decays as $\omega_{q\eta}/J_{1}^0$ \cite{Uhrig1998}. For now though, we set $g_{\vect{q}} = 1$.  Next, we average over transformed phonon degrees of freedom \cite{Kennes2017}, leading to an effective spin Hamiltonian, $H_{\mathrm{eff}}$. Using the cancellation provided by Eq. (\ref{eq:cancelation}) and discarding vanishing phonon averages, $H_{\mathrm{eff}}$ is
\begin{equation}\label{eq:Heff}
    \!H_{\mathrm{eff}}\!=\!\langle \tilde{H} \rangle_{H_p}
    \!\approx\! H_s\!+\! \frac{1}{2}\langle\left[S,H_{sp}\right]\rangle_{H_p}\!\!+\!\frac{1}{2}\langle\left[S,\left[S,H_s\right]\right]\rangle_{H_p}\!
\end{equation}
to second order in $\lambda$.  The three terms on the right hand side will be of order $\left(J_1^{(0)}\right)^m$, $m \leq 3$.

We can further separate these terms into those independent of the phonon distribution, and thus present at zero temperature in equilibrium ($H_{i,0}$) and those dependent on the phonon distribution ($H_{i,n}$).  It is useful to treat the second and third terms in Eq. (\ref{eq:Heff}) separately, as they involve four and six sites, respectively.  We define,
\begin{align}\label{eq:Heffterms}
    \frac{1}{2}\langle\left[S,H_{sp}\right]\rangle_{H_p}&=H_{1,0}+H_{1,n}\cr
    \frac{1}{2}\langle\left[S,\left[S,H_s\right]\right]\rangle_{H_p}&=H_{2,0}+H_{2,n},
\end{align}
where the distribution independent terms are:
\begin{align}
    H_{1,0}=&-\frac{1}{2N}\!\!\sum_{\substack{\vect{q}\eta,\\ i \zeta_i, j \zeta_j}}\!\!\alpha_{\vect{q}\eta}^2e^{i\vect{q}\cdot \left(\vect{r}_i-\vect{r}_j\right)} \left(T_{\zeta_i\zeta_j}^{\vect{q}\eta}+T_{\zeta_i\zeta_j}^{*-\vect{q}\eta}\right) \cr & \times \left(\vect{S}_i \cdot \vect{S}_{\zeta_i}\right) \left(\vect{S}_j \cdot \vect{S}_{\zeta_j}\right), \label{eq:HeffEterms1}\\
    H_{2,0}=&-\frac{J_1^{(0)}}{2N}\!\!\sum_{\substack{\vect{q}\eta, i \zeta_i,\\ j \zeta_j,k \zeta_k}}\!\!\alpha_{\vect{q}\eta}^2\frac{J_{1}^{(0)}}{\omega_{\vect{q}\eta}}e^{i\vect{q}\cdot \left(\vect{r}_i-\vect{r}_j\right)}\left(T_{\zeta_i\zeta_j}^{\vect{q}\eta}-T_{\zeta_i\zeta_j}^{*-\vect{q}\eta}\right)\cr
    &\times\left(\vect{S}_i \cdot \vect{S}_{\zeta_i}\right)\left[\vect{S}_j \cdot \vect{S}_{\zeta_j},\vect{S}_k \cdot \vect{S}_{\zeta_k}\right],\label{eq:HeffEterms2}
\end{align}
where
\begin{align}\label{eq:Tmatdef}
T_{\zeta_i\zeta_j}^{\vect{q}\eta}=f^{*i\zeta_i}_{\vect{q}\eta}f^{j\zeta_j}_{\vect{q}\eta},
\end{align}
the dimensionless coupling constant $\alpha_{\vect{q}\eta}$ is given by:
\begin{align}\label{eq:alphadef}
    \alpha^2_{\vect{q}\eta}=\lambda^2\frac{1}{2M \omega_{\vect{q}\eta} }\frac{J_{1}^{(0)}}{\omega_{\vect{q}\eta}},
\end{align}
while $\zeta_i$ labels the nearest neighbors of spin $i$. These are the effective spin-spin interactions induced by zero-point lattice fluctuations. The zero-point fluctuation induced spin-spin interactions have been a topic of extensive research, as they are responsible for the spin-Peierls transition \cite{Kuboki1987, Uhrig1998, Weisse1999, Zhang2008}. Here, we instead focus on the distribution-dependent terms as they capture phonon pumping effects.

The phonon distribution, $n_{\vect{q}\eta}=\langle b_{\vect{q}\eta}^\dagger b_{\vect{q}\eta} \rangle_{H_p}$ dependent terms have the following structure:
\begin{align}
    H_{1,n}=&\frac{J_1^{(0)}}{ 2N}\!\!\sum_{\substack{\vect{q}\eta,\\ i \zeta_i, j \zeta_j}}\!\!\alpha_{\vect{q}\eta}^2e^{i\vect{q}\cdot \left(\vect{r}_i-\vect{r}_j\right)}\label{eq:Heffpterms1}\\
    &\times\left[T_{\zeta_i\zeta_j}^{\vect{q}\eta}(1+n_{\vect{q}\eta})-T_{\zeta_i\zeta_j}^{*-\vect{q}\eta}n_{-\vect{q}\eta}\right]\cr
    &\times\left[\vect{S}_i \cdot \vect{S}_{\zeta_i},\vect{S}_j \cdot \vect{S}_{\zeta_j}\right],\cr 
    H_{2,n}=&-\frac{J_1^{(0)}}{2N}\!\!\sum_{\substack{\vect{q}\eta, i \zeta_i,\\ j \zeta_j,k \zeta_k}}\!\!\alpha_{\vect{q}\eta}^2\frac{J_{1}^{(0)}}{\omega_{\vect{q}\eta}}e^{i\vect{q}\cdot \left(\vect{r}_i-\vect{r}_j\right)}\cr
    &\times\left[T_{\zeta_i\zeta_j}^{\vect{q}\eta}(1+n_{\vect{q}\eta})+T_{\zeta_i\zeta_j}^{*-\vect{q}\eta}n_{-\vect{q}\eta}\right]\cr
    &\times\left[\vect{S}_i \cdot \vect{S}_{\zeta_i},\left[\vect{S}_j \cdot \vect{S}_{\zeta_j},\vect{S}_k \cdot \vect{S}_{\zeta_k}\right]\right].\label{eq:Heffpterms2}
\end{align}
These contain contributions from both stimulated phonon emission ($1+n_{\mathbf{q}}$) and absorption ($n_{\mathbf{q}}$). While prior work has noted the possibility of thermal phonon populations affecting the magnetic interactions \cite{Uhrig1998, Weisse1999}, we are particularly interested in the distinct potential effects of \emph{non-equilibrium} phonon populations.

For general spins, the generated magnetic couplings will usually involve three ($H_1$) or four ($H_2$) spin terms, due to the nested commutator structure. The expressions are significantly simplified for $S = 1/2$, where any four spin interactions with a repeated site will reduce to a sum of two and three spin terms,
\begin{equation}\label{eq:12algebra}
    (\vect{S}_i \cdot \vect{S}_j)(\vect{S}_i \cdot \vect{S}_k)=\vect{S}_j \cdot \vect{S}_k +i  \vect{S}_i \cdot \left( \vect{S}_j \times \vect{S}_k\right).
\end{equation}
Overall, the $S=1/2$ case will generate: two-spin exchange interactions, three spin chirality terms, and a variety of four spin terms including ring exchange.  Once we sum over all $(ijkl)$, we find that the four spin terms cancel to exactly zero in all of the relatively simple models studied in this paper.  However, they are not forbidden, and we expect that more complicated or less symmetric models will be able to generate four spin terms. The formalism we present can straightforwardly be generalized to include further neighbor exchange terms in Eqs. (\ref{eq:Hs}) and (\ref{eq:Hsp}), as well as to treat higher spins, where biquadratic and other interactions will be generated, makingthe calculations significantly more tedious.

Now we focus on $H_{1,n}$ and  $H_{2,n}$, which involve nonequilibrium phonon distributions.  These are most straightforwardly achievable experimentally by driving IR active optical phonons with light \cite{Nova2017, Cartella2018, Disa2020, Afanasiev2021}.  The main simplification is then that these nonequilibrium phonon distributions are non-zero only around the Brillouin zone center ($n_{\mathbf{q}=0,\eta}=n_{
\eta}$), and the resulting effective spin interactions can be separated into symmetric and antisymmetric parts arising from $H_{2,n}$ and $H_{1,n}$, respectively,
\begin{align}
    H_{A}=& J_1^{(0)}\!\!\sum_{\substack{\eta, i \zeta_i\\j \zeta_j}}\!\alpha_\eta^2 n_{\eta}\Im T_{\zeta_i \zeta_j}^{\eta} i\left[\vect{S}_i \cdot \vect{S}_{\zeta_i},\vect{S}_j \cdot \vect{S}_{\zeta_j}\right], \label{eq:Aterm}\\
    H_{S}=&-\!J_1^{(0)}\!\!\!\sum_{\substack{\eta, i \zeta_i,\\ j \zeta_j,k \zeta_k}}\!\!\alpha_\eta^2 n_{\eta}\frac{J_{1}^{(0)}}{\omega_0}\Re T_{\zeta_i \zeta_j}^{\eta}\cr
    & \times\left[\vect{S}_i \cdot \vect{S}_{\zeta_i},\left[\vect{S}_j \cdot \vect{S}_{\zeta_j},\vect{S}_k \cdot \vect{S}_{\zeta_k}\right]\right]. \label{eq:Sterm}
\end{align}

The symmetric term, $H_S$ is present for any driven phonon mode with linear spin-phonon coupling, and generates magnetic exchange interactions.  The antisymmetric term, $H_A$ can only be driven where the phonon modes are degenerate, as non-degenerate modes have real eigenvectors, $f_\eta$, due to the symmetric nature of the dynamical matrix.  For doubly degenerate modes, there is a natural choice of left and right circularly polarized phonon modes, $f_{L,R}$ for which $\Im(T_R)=-\Im(T_L)$.  If one can selectively pump one such polarization (see Sec. \ref{sec:SPpump} for details), the antisymmetric term will generate three-spin chiral interactions with a linear dependence on the drive asymmetry,
\begin{equation}\label{eq:betadef}
    \beta=\frac{n_{R}-n_{L}}{n_{R}+n_{L}}.
\end{equation} 
In the rest of the paper, without loss of generality, we consider a single pumped phonon mode, adopting the notation $\alpha=\alpha_{\eta}$ and $n_{ph} = n_{\eta}$, with $n_{ph}=n_{R}+n_{L}$ for polarized pumping.

We note the critical role that quantum fluctuations play in the mechanism of spin interaction tuning through phonon pumping we consider here. The more mutually non-commuting terms the Hamiltonian has, the larger the prefactor in the pumped terms stemming from Eqs. (\ref{eq:Sterm}-\ref{eq:Aterm}) will be, and the more diverse the generated spin couplings will be. The relation is not direct, as the degree of non-commutativity and the underlying lattice connectivity together define the strength of spin-phonon pumping, as presented in detail on 2D lattice examples in Sec. \ref{sec:Results}. Still, a purely classical spin Hamiltonian leads to zero spin-phonon pumping in the leading order, and thus, spin quantum fluctuations are the ``resource" necessary for spin-phonon pumping.

Finally, we consider the validity of the perturbation expansion based on the Lang-Firsov transformation, which is expected to be valid as long as the underlying coupling constants are sufficiently small, and the phonons remain in the diabatic regime. There are two independent conditions to be satisfied. First, the coupling constant condition for the phonon distribution dependent terms requires both $\alpha^2 \ll 1$ and $\alpha^2 n_{ph} \ll 1$, as the next order in the perturbation expansion will generically have both $\mathcal{O}(\alpha^4 n_{ph})$ and $\mathcal{O}(\alpha^4 n_{ph}^2)$ terms. \cite{Weisse1999} Anticipating a typical pumping experiment with $n_{ph}\gtrsim 1$, $\alpha^2 n_{ph} \ll  1$ covers both conditions. Secondly, $J_1^{(0)}/\omega_0 < 1$ is required to remain in the diabatic regime \cite{Weisse1999, Bursill1999}.  While we focus on the diabatic case here, we note that the adiabatic case can also be treated via a variational version of Lang-Firsov transformation [$g_{\vect q} \neq 1$ in Eq. (\ref{eq:LFtransf})]. 
 The variational parameter $g$ must be determined self-consistently, and will in principle depend on the details of spin-dynamics. An idea of the limits of validity can be obtained from 1D Hamiltonians \cite{Weisse1999} where the variational parameter was found to vary smoothly from $g \rightarrow 0$ in the adiabatic limit ($\omega_0/J_1^{(0)} \rightarrow 0$) to close to 1 for $J_1^{(0)}/\omega_0\ \rightarrow 0$, with a typical crossover at $J_1^{(0)}/\omega_0 \sim 1$. 
 The small $g$ in the adiabatic limit indicates a diminished effectiveness of the spin-phonon coupling, making the diabatic case considered here likely to have the strongest spin-phonon effects. Higher order terms or even higher order spin-phonon couplings \cite{Yarmohammadi2023, Yarmohammadi2024b} may be relevant for more quantitative predictions, particularly in materials with strong spin-phonon couplings.  Generically, these terms will modify the induced exchange couplings we discuss here, as well as induce additional terms involving more neighbors and more spins. We also note that, within our Hamiltonian, the phonon-induced spin-relaxation time is parametrically long, scaling as $\propto 1/\alpha^2$. This relaxation is therefore typically slower than the intrinsic spin-correlation timescale ($\propto 1/J$). Moreover, because the lattice displacements are driven at a frequency $\omega \gg \max\{J,\lambda\}$, the dynamics are accurately captured by the effective average Hamiltonian of Eq.~(\ref{eq:Heff}) as Floquet pre-thermalization~\cite{Abanin2017} ensures that both heating and drive-induced dissipation are exponentially suppressed on timescales $\propto \exp\left[\frac{\omega}{\max(J,\alpha)}\right]$. As a result, the dominant dissipative channels like disorder, magnon-magnon, or spin-phonon relaxation remain essentially the same as in the undriven system. Whether these intrinsic dissipative processes are weak enough to allow correlated spin states to emerge is therefore determined by the undriven case -- if coherent low-energy spin excitations are observable without the drive, they should likewise be observable in the driven setting considered here.

Exact numerical studies of 1D spin-phonon models show that the Lang-Firsov perturbation expansion has significant predictive power. In particular, the predicted critical spin-phonon coupling for the dimerization transition in a $J_1-J_2$ $S=1/2$ chain agrees well with density matrix renormalization group  \cite{Bursill1999, Hager2007} and quantum Monte Carlo results \cite{Sandvik1999}. The agreement is unusually good, particularly in the diabatic regime where the critical couplings match even for dimensionless coupling constants of the order of $\sim 0.3$. As expected, the perturbation expansion is less successful in the adiabatic regime, even with the variational approach. In this limit, the description of physics in terms of modified exchange interactions itself can fail, as shown in 2D quantum Monte Carlo simulations of the adiabatic spin-phonon model on the honeycomb lattice \cite{Weber2021}.

\section{Experimental feasibility\label{sec:SPpump}}

While it is possible to generate these effective spin Hamiltonians with thermal phonon occupations, these are typically not large enough to induce sizable couplings, and the high temperatures can disorder the spins. A more promising route is to pump a non-equilibrium phonon distribution \cite{Nova2017, Cartella2018, Disa2020, Afanasiev2021}. Here, we see that the currently accessible experimental couplings are large enough to drive significant magnetic interaction changes, and that phonon polarization can be an additional tuning parameter to generate chiral interactions.

The most direct experimental implementation is driving IR active phonon modes by mid-IR light. This drive will generically induce a non-equilibrium phonon density matrix whose off-diagonal components quickly decay due to phonon-phonon and electron-phonon coupling \cite{Kennes2017}, leaving the diagonal components corresponding to phonon occupancy, $n_{ph}$, as captured by the Lang-Firsov approach above \cite{Kennes2017}. Driving significant changes to the magnetic exchange interactions typically requires a sizable $\alpha^2 n_{ph}$, obtained by driving a relatively large $n_{ph}$. As an example of current experimental capabilities, we discuss phonon pumping experiments on DyFeO$_3$ \cite{Afanasiev2021} for the IR active $B_{1u}$ mode \cite{Juraschek2017}. Depending on the ab initio extracted values of $\lambda$, which are highly mode and bond dependent, $\alpha^2 n_{ph}$ can be in the range of $0.1-10$. Even with the fairly small $J_1^{(0)}/ \omega_0$ ratio in this material, there can be a sizable $\alpha^2 n_{ph} J_1^{(0)}/ \omega_0$ driving coefficient in the range of $0.01-1$, with the higher end beyond the validity of perturbation theory.  We will show that this range can drive significant interaction changes and drive materials into spin liquid regimes.  These large couplings are achieved mostly through large phonon occupancies, estimated to be $n_{ph} \gtrsim 20$, also in line with a $K_3 C_{60}$ phonon driving experiment \cite{Mitrano2016, Kennes2017}. These large occupations can compensate for smaller $\lambda$'s and $J_1^{(0)}/ \omega_0$ ratios, and allow for spin-correlation time scales that are much faster than longer spin-relaxation time scales; in DyFeO$_3$ these are $<5ps$ and $100ps$, respectively \cite{Juraschek2017}.

In addition to the relatively large coupling constants, an additional advantage of optical phonon pumping lies in the suppressed heating of the electronic degrees of freedom. This suppression is in contrast with Floquet engineering of magnetism in Mott insulators, where heating effects \cite{Claassen2017, Alessio2014} may be significant and require sub-gap light that limits the tunability \cite{Quito2021}. Here, the relevant optical phonon frequencies are naturally in the sub-gap range and thus produce negligible electronic heating, while still being accessible to IR light. The main source of electronic heating in the spin-phonon pumping is indirect heating through the electron-phonon interaction, which is off-resonance and thus less problematic.

Direct pumping is possible only for IR active phonons, but Raman active phonons can still be indirectly pumped. As heating is a concern, Raman pumping through electronic levels should be avoided, but it is also possible do non-linear Raman pumping using IR light \cite{Forst2011, Afanasiev2021, Zeng2023}. The coupling between Raman and IR active modes stems from anharmonic phonon terms and can be used to drive the Raman mode population \cite{Subedi2014}. The typical Raman-active phonon occupancies achieved with this method are two orders of magnitude below the IR case \cite{Afanasiev2021}, making significant magnetic exchange tuning beyond the current experimental capabilities. However, improvements in laser power could lead to significant gains in non-linear Raman pumping, as $n_{ph} \propto I^2$, where $I$ is the intensity of light, contrasted with $n_{ph} \propto I$ for direct IR active pumping.  Also, while Raman active modes are necessary for some simple models we consider below, real materials typically have many IR active modes to choose from, making Raman-only active pumping unnecessary.

Driving chiral interactions requires pumping left or right circularly polarized phonon distributions \cite{Fransson2023, Luo2023, Ohe2024, Wu2024}, with non-zero drive asymmetry, $\beta$.  These are induced only when the driving field breaks time-reversal, which can be done using circularly polarized light.  For IR active doubly degenerate modes, circular light will directly induce the desired phonon polarization, making chiral pumping readily experimentally accessible.  Raman-active modes are more challenging; in the proposed non-linear pumping schemes for Raman modes \cite{Subedi2014}, pumping IR phonons will typically not induce polarization for the degenerate Raman modes. The interplay of both circularly polarized pump and coupling to IR active modes might allow for non-linear polarized Raman pumping, however, it is likely beneficial to chose materials with IR active degenerate modes amenable to direct polarized pumping.  
Generic elliptical polarizations will likely drive spatially anisotropic spin couplings if the phonon mode is doubly degenerate, which should be treatable with an extension of this method, although we have not considered that here; unpolarized light will pump both modes incoherently, with $\beta = 0$.
By contrast, engineering the light polarization profile can have a more direct effect when Floquet engineering the electronic degrees of freedom  \cite{Quito2021, Quito2021b}.


\section{Results on 2D lattices\label{sec:Results}}

Having described the general mechanism for non-equilibrium phonon pumping to affect the magnetic interactions, we now apply it to several simple 2D lattices and spin Hamiltonians, chosen to showcase different generic features. Direct exchange is considered on the honeycomb and kagom\'{e} lattices (Appendix \ref{sec:kagome}), and superexchange interactions are considered on the square and triangular lattices (Appendix \ref{sec:Triangular}).  In all cases, we restrict ourselves to nearest-neighbor models, which capture the effects of phonon mode selection and phonon polarization on different exchange mechanisms.  The techniques generalize straightforwardly to further neighbor interactions.  Finally, we explore the interplay of magnetic anisotropy and spin-phonon pumping on the honeycomb lattice with both XYZ, pure Kitaev, and Kitaev-Heisenberg interactions, where we show that spin-phonon pumping can be used to increase the anisotropy and favor Kitaev terms.

\subsection{Direct exchange interactions\label{sec:HCheis}}


We first describe the simpler case of direct exchange interactions in order to present the mechanism of spin-phonon pumping. In particular, we consider the nearest-neighbor Heisenberg model on the honeycomb lattice. The mono-atomic honeycomb lattice provides a particularly simple example, as it hosts one spin-phonon active optical mode, while giving qualitatively representative results for more complex lattices. The more complex example of a kagome lattice with multiple spin-phonon active modes is treated in Appendix \ref{sec:kagome}.

\begin{figure}[!htb]
\includegraphics[width=1.0\columnwidth]{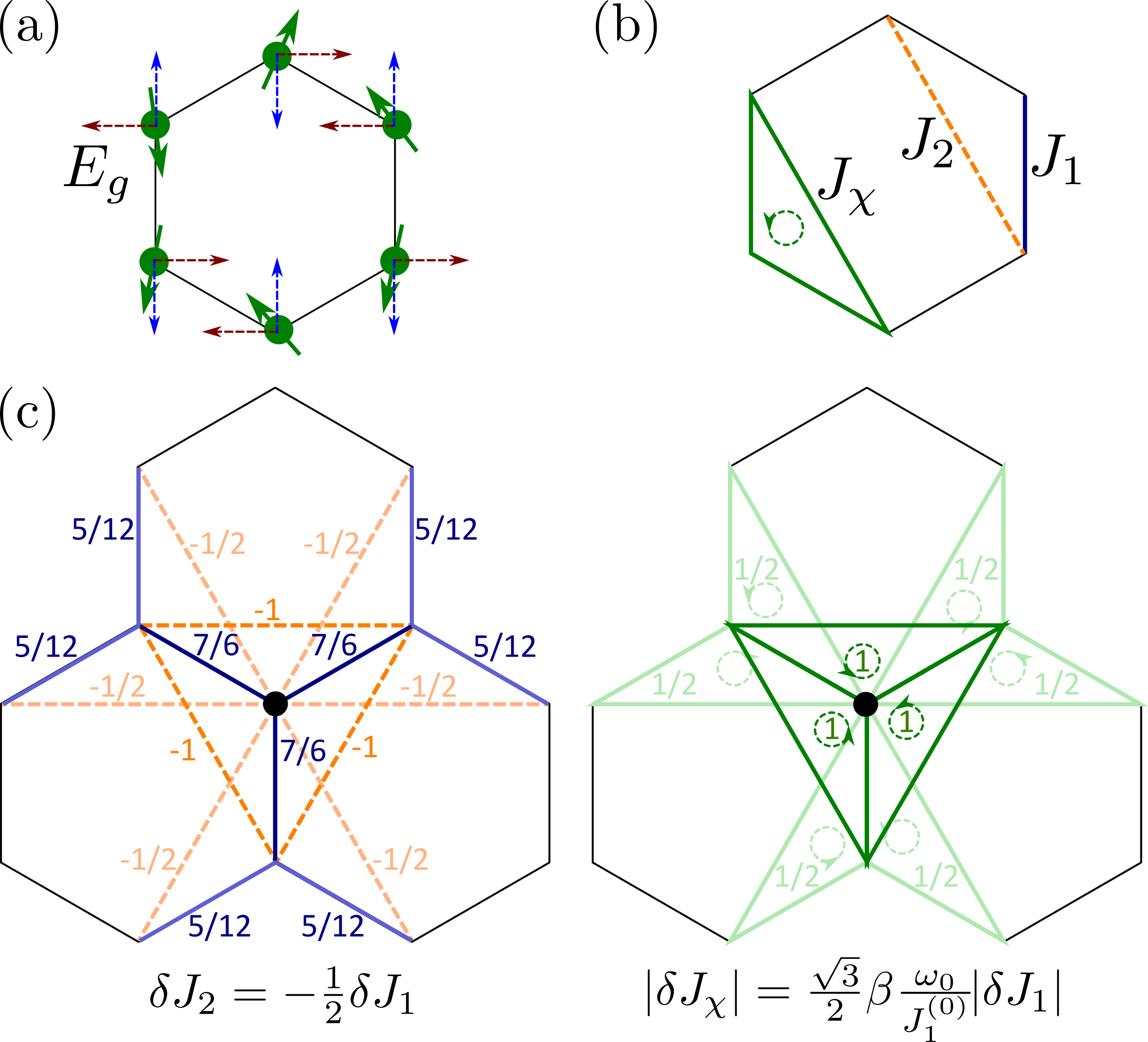}
\caption{Honeycomb phonons and exchange couplings. (a) The optical phonon mode, $E_g$ couples linearly to the spins. (b) Exchange couplings generated by $E_g$ phonon pumping for the nearest neighbour Heisenberg model. (c) $S=1/2$ exchange (left) and chiral (right) couplings arising from the center site ($i=0$) of Eq. (\ref{eq:Aterm}-\ref{eq:Sterm}). Exchange (chiral) interactions have units of $J_1^{(0)} \alpha^2 n_{ph} \frac{J_1^{(0)}}{\omega_0}$ ($\frac{\sqrt{3}}{2} J_1^{(0)} \alpha^2 n_{ph} \beta$). Using translational invariance leads to Eq. (\ref{eq:Hccouplings}).  \label{fig:Fig2_honeycomb_couplings}
}
\vspace{-0.cm}
\end{figure}

The honeycomb lattice has two atoms in the unit cell, and hosts three optical phonon modes: the non-degenerate $A_{1g}$ and the doubly degenerate $E_g$. We consider direct exchange interactions, and the relevant phonon mode displacement must have a non-zero projection onto the bond direction, as in Eq. (\ref{eq:r0HC}), which means only the $E_g$ mode is spin-phonon active. Anticipating the possibility of polarized pumping, as discussed in Sec. \ref{sec:SPpump}, we write the degenerate mode in the $R/L$ basis as:
\begin{equation}\label{eq:HCmodes}
    \hat{\epsilon}_{E_g,R}=\frac{1}{\sqrt{2}}(\hat{x}+i\hat{y}), \qquad \hat{\epsilon}_{E_g,L}= \hat{\epsilon}_{E_g,R}^*.
\end{equation}
These are the displacements for the first atom in the unit cell, with the second having the opposite displacement, as shown in Fig. \ref{fig:Fig2_honeycomb_couplings}(a). 

From the phonon mode displacements, bond directions, and the geometric positions of the atoms, we construct the $T$ given by Eq. (\ref{eq:Tmatdef}). When considering the nearest neighbor spin-phonon coupling, the relevant part of $T$ is indexed by the three nearest-neighbor bonds:
\begin{equation}\label{eq:TmatHC}
    T^{E_g,R}=\begin{pmatrix}
1 & \xi & \xi^* \\
\xi^* & 1 & \xi \\
\xi & \xi^* & 1 \\
\end{pmatrix}, \qquad T^{E_g,L}= T^{*E_g,R},
\end{equation}
where $\xi=e^{2\pi i/3}$.

We now evaluate the commutators of Eq. (\ref{eq:Aterm}) and (\ref{eq:Sterm}). The evaluation of the terms, while straightforward, is very tedious, since the symmetric terms contain six spin nested commutators, with three lattice and three nearest-neighbor sums.  We used a computer algebra package, DiracQ \cite{Wright2013} to do this evaluation.  Note that the six spin nested commutators are only non-zero when two indices are shared between the three bond terms. In practice, we fix $i=0$ as the center site.  Now all other involved sites are at most three bond lengths away, vastly reducing the terms in the sum. We then apply lattice translations to obtain the final effective spin couplings.

As a general rule, the nested commutator structure of the phonon distribution interaction terms leads to two-, three-, and four-spin  interactions in $S=1/2$ systems, as shown in Eq. (\ref{eq:12algebra}). The nested commutator structure combined with the nearest-neighbor initial spin interactions leads to a ``three-bond rule'', which means that the generated effective spin terms are between sites connected by at most three bonds. On the honeycomb lattice, this results in modified nearest-neighbor interactions ($\delta J_1$) and the generation of effective next-nearest-neighbor exchange ($J_2$) and chiral coupling ($J_{\chi}$) terms, as shown in Fig. \ref{fig:Fig2_honeycomb_couplings}(b). In addition, four-spin terms of the form $(\vect{S}_{i}\cdot \vect{S}_{j})(\vect{S}_{k}\cdot \vect{S}_{l})$ with at most one bond length between the $(ij)$ and $(kl)$ bonds are allowed, however we omit them from the further discussion as the explicit calculation shows that their magnitude vanishes. For the non-vanishing interactions, we show the intermediate result of evaluating Eq. (\ref{eq:Aterm}) and (\ref{eq:Sterm}) for $i=0$ in Fig. \ref{fig:Fig2_honeycomb_couplings}(c), which gives a sense of which bonds contribute. The intermediate results explicitly show the origin of the three bond rule and the limited number of generated interactions. The full calculation gives the spin interactions:
\begin{align}\label{eq:Hccouplings}
    J_1=&J_1^{(0)}+\delta J_1=J_1^{(0)}\left(1-6\alpha^2 n_{ph}\frac{J_1^{(0)}}{\omega_0}\right)\cr
    J_2=&\delta J_2=3 J_1^{(0)}\alpha^2 n_{ph}\frac{J_1^{(0)}}{\omega_0}\cr
    J_{\chi}=&\delta J_{\chi}=3\sqrt{3}J_1^{(0)}\alpha^2 n_{ph} \beta
\end{align}

These expressions show several general features of spin-phonon pumped interactions.  The exchange interactions, $\delta J_1$ and $J_2$ arise from the symmetric term, Eq. (\ref{eq:Sterm}), and are proportional to $\alpha^2 n_{ph}\frac{J_1^{(0)}}{\omega_0}$; these have opposite signs, with $J_1$ reduced by phonon pumping. This result holds on many lattices, as we shall show, and stems from the different commutator structures driving $\delta J_1$ and $J_2$.  The chiral interaction is generated exclusively from the asymmetric term,  Eq. (\ref{eq:Aterm}), and is thus proportional to $\alpha^2 n_{ph}$ and only non-zero with asymmetric drive polarization, $\beta$.  The honeycomb structure affects the numerical prefactors and restricts the allowed changes to the interactions. The behavior described above is valid for both signs of $J_1^{(0)}$: in the (anti)ferromagnetic case, (anti)ferromagnetic next neighbor interactions are driven, and $|J_1|$  decreases. Here, we focus on the antiferromagnetic case, as the pure ferromagnetic case does not induce frustration.  

\begin{figure}[!htb]
\includegraphics[width=1.0\columnwidth]{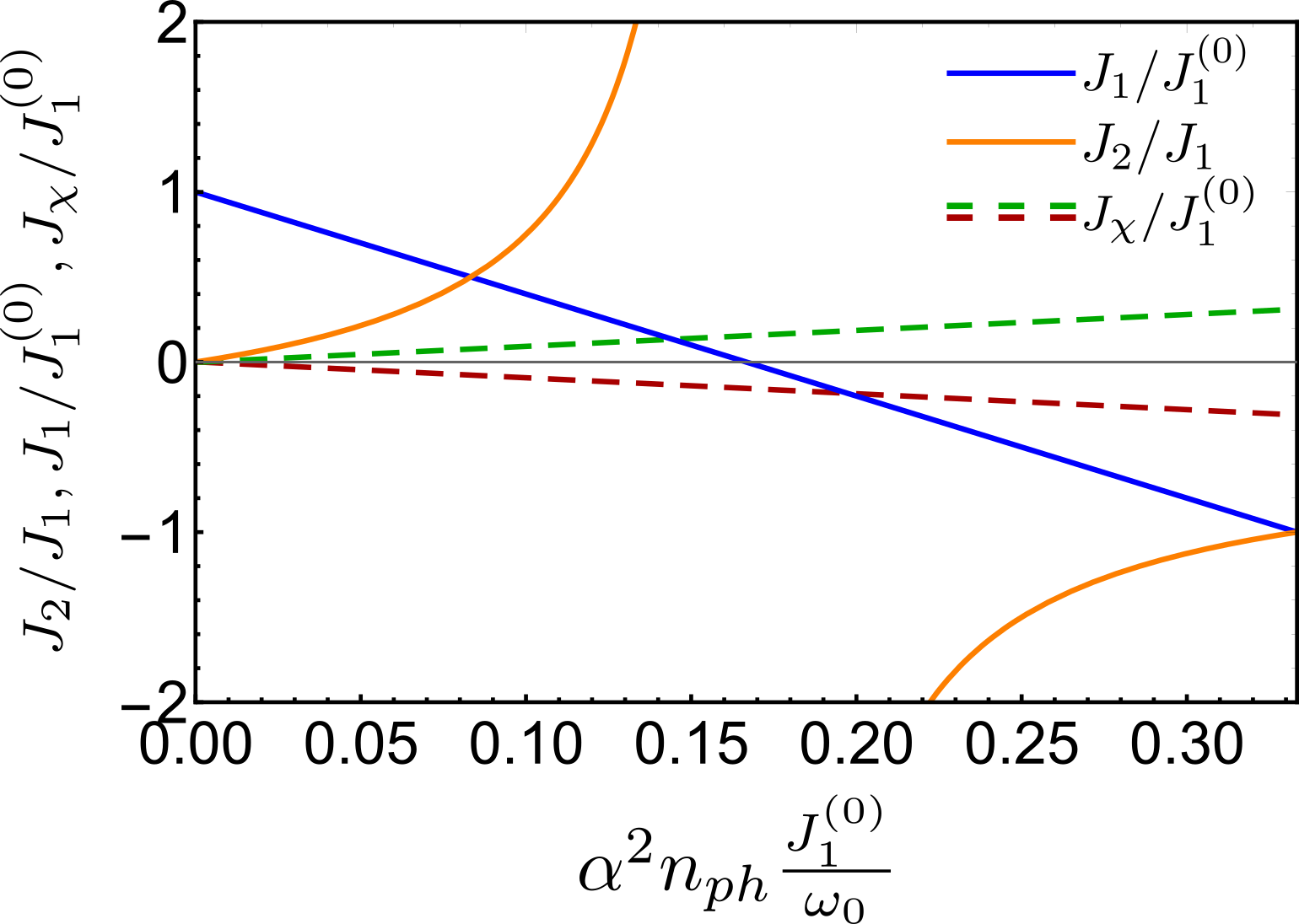}
\caption{Spin-phonon pumping of the honeycomb Heisenberg model with $E_g$ phonon modes.  Nearest,  next-nearest neighbor, and chiral exchange couplings as a function of $\alpha^2 n_{ph}J_1^{(0)}/\omega_0$; note that $J_1$ vanishes for $\alpha^2 n_{ph}J_1^{(0)}/\omega_0 \approx .15$, leading to a divergence in $J_2/J_1$. A deconfined critical point or spin-liquid region is predicted for the $J_1-J_2$ honeycomb model at $J_2/J_1\sim .22$ \cite{Clark2011, Albuquerque2011, Ganesh2013}, here found with $\alpha^2 n_{ph}J_1^{(0)}/\omega_0 \sim .1$. Induced chiral fields ($J_\chi$) are plotted as a function of $\alpha^2 n_{ph}$ for $\omega_0/J_1^{(0)}=1/2$ and two opposite phonon polarizations, $\beta = 1$ (green, dashed) and $\beta=-1$ (red, dashed). A chiral spin liquid is predicted for $J_\chi/J_1 \sim .25$ \cite{Hickey2016}, here achieved for $\alpha^2 n_{ph} \sim .12$ and full polarization.}\label{fig:Fig3_honeycomb_res}
\vspace{-0.cm}
\end{figure}


The $J_1-J_2-J_\chi$ honeycomb Heisenberg model has rich physics: increasing $J_2$ stabilizes a plaquette valence bond solid above $J_2/J_1 = .22$, with a potential intermediate spin liquid or deconfined critical point \cite{Clark2011, Albuquerque2011, Ganesh2013}, while $J_\chi$ can stabilize a chiral spin liquid for $J_\chi/J_1 >.25$ \cite{Hickey2016}.  Both regions are potentially accessible by spin-phonon pumping.  To showcase this, we plot $J_2/J_1$ and $J_{\chi}/J_1$ as a function of relevant spin-phonon couplings in Fig. \ref{fig:Fig3_honeycomb_res}. Reaching sizeable $J_2/J_1$ is plausible for $\alpha^2 n_{ph}J_1^{(0)}/\omega_0$ in the range of $0.01-1$, which is accessible in current experiments; $J_2/J_1\sim 0.2$ requires only $\alpha^2 n_{ph}J_1^{(0)}/\omega_0\sim 0.1$, with $J_1 \approx 0.4 J_1^{(0)}$ still significant. $J_{\chi}/J_1$ can be even larger, as it is proportional to $\alpha^2 n_{ph}$, and typically $J_1^{(0)}/\omega_0<1$ in the regime of interest, although it is unclear what degree of phonon polarization, $\beta$ is experimentally achievable.

\subsection{Superexchange interactions} \label{sec:SEpaths}

In insulating magnetic materials, most interactions are generated by superexchange rather than direct exchange, typically involving virtual hopping through an intermediate ligand ion.  In this section, we consider simple superexchange models on the square lattice and derive general rules governing spin-phonon pumping of superexchange spin interactions. The direct extension to the triangular lattice is presented in Appendix \ref{sec:Triangular}.

\begin{figure}[!htb]
\includegraphics[width=1.0\columnwidth]{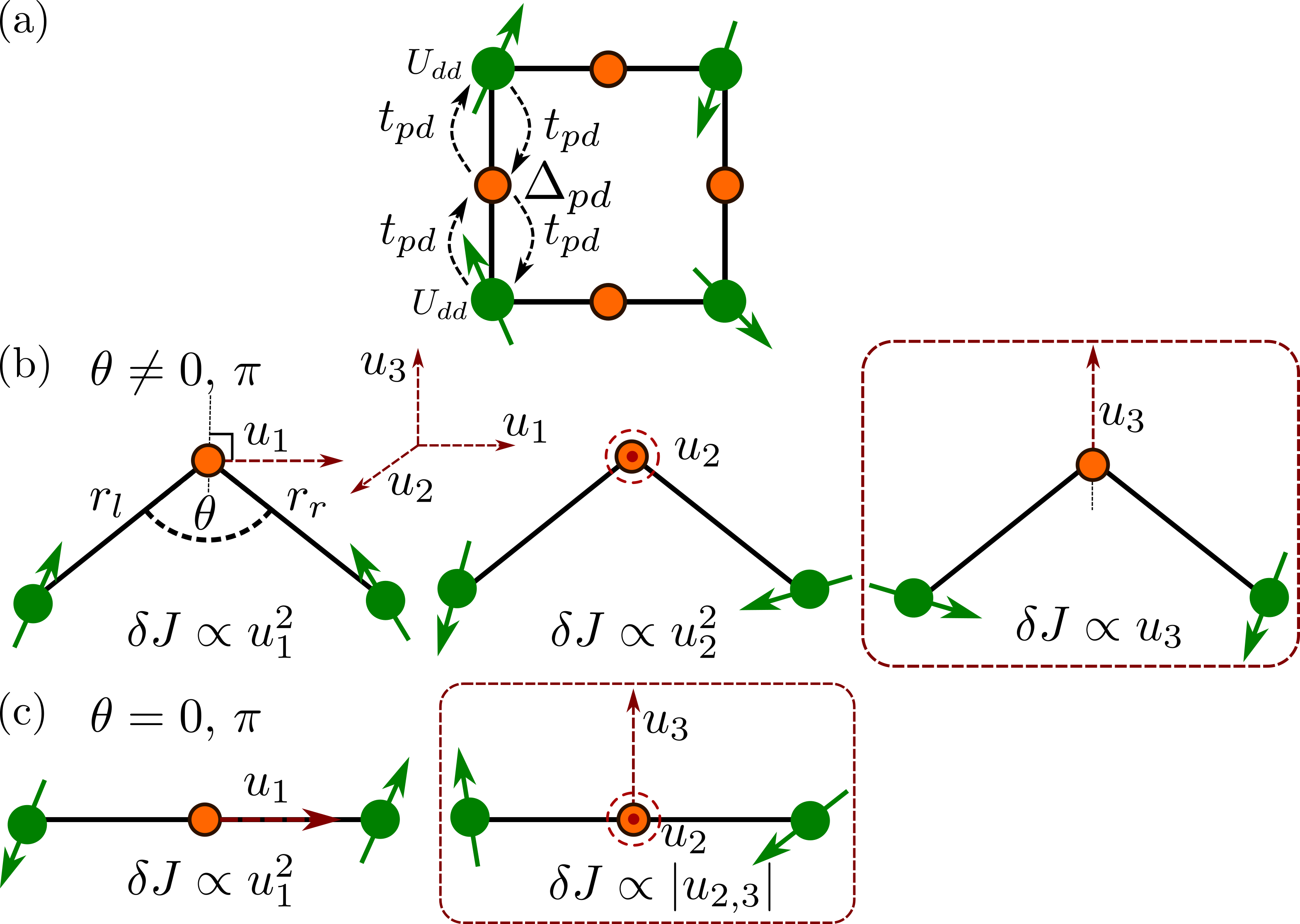}
\caption{Superexchange and linear spin-phonon coupling. (a) Superexchange paths typical of CuO$_2$ planes. The exchange interaction is mediated by electron hopping through the central anion resulting in $J_{1} \propto t_{pd}^4/\Delta_{pd}^3$. $J_1$ is sensitive to both the total length of the metal-anion bonds, as well as the angles involved. The phonon pumping has a linear effect on $J_{1}$ only if the right and left parts of the superexchange path add constructively, which leads to geometric constraints on the relevant phonon modes. (b) For a non-collinear superexchange path, linear spin-phonon coupling is present only for phonon displacements perpendicular to the bond, $u_3$. (c) For collinear superexchange paths, any orthogonal phonon displacement leads to a linear spin-phonon coupling. \label{fig:Fig4_superexchange}}
\vspace{-0.cm}
\end{figure}

Motivated by CuO$_2$ planes, we consider an edge-stuffed square lattice model with spins on the vertices and superexchange mediating ligands on the edges, shown in Fig. \ref{fig:Fig4_superexchange}(a).  The nearest neighbor magnetic exchange is generated in fourth order perturbation theory \cite{Zhang1988}:
\begin{equation}\label{eq:cuprateJ}
    J_1 = \frac{2t_{pd}^4}{\Delta_{pd}^3}\left(1+\frac{U_{dd}}{2\Delta_{pd}}\right),
\end{equation}
where $t_{pd}$ is the metal-anion hopping, $U_{dd}$ is the $d-$electron Hubbard interaction and $\Delta_{pd}$ is the charge transfer energy.
We first construct a spin-phonon coupling model for generic superexchange paths before applying it to the square lattice. $J_{1}$ depends on both the path length and angle \cite{Rosch2004,Rocquefelte2012}. A given spin-phonon deformation couples linearly via an effective coupling,
\begin{equation}\label{eq:lamSE}
\lambda=\frac{1}{J_1^{(0)}}\left(\frac{dJ_1(\vect{r})}{d (r_l+r_r)}+\frac{dJ_1(\vect{r})}{r d\theta}\right),
\end{equation}
where $r_{l(r)}$ are changes in the bond distances for the left (right) parts of the superexchange path in Fig. \ref{fig:Fig4_superexchange}(b); we keep these individual terms, even though they add, as there can be non-trivial displacements with $\delta(r_{l}+r_{l})=0$; it is clear that these will not generate linear terms.

We take a typical symmetric superexchange path connecting two equivalent metal ions, with metal-anion-metal angle, $\theta$. In this case, the superexchange path distortion can be completely described in terms of the relative displacement of the anion, $\vect{u}$. We decompose the displacement into the three physically relevant directions shown in Fig. \ref{fig:Fig4_superexchange}(b): $u_1$, parallel to the metal-metal direction; $u_2$ orthogonal to the plane of the path; and $u_3$ orthogonal to both $u_1$ and $u_2$. Out of the three, only the $u_3$ direction generically leads to linear spin-phonon coupling.  $u_3$ is defined by the unit vector:
\begin{equation}\label{eq:r0SE}
    \hat{u}_3^{ij}=-\frac{\vect{r}_l+\vect{r}_r}{\left|\vect{r}_l+\vect{r}_r\right|}=-\frac{\vect{r}_i+\vect{r}_j-2\vect{r}_{A}^{ij}}{\left|\vect{r}_i+\vect{r}_j-2\vect{r}_{A}^{ij}\right|},
\end{equation}
where $\vect{r}_A$ and $\vect{r}_{i,j}$ are the anion and metal locations, respectively.  Linear spin phonon coupling arises as
\begin{equation}\label{eq:deltarth}
    \delta (r_l+r_r) \propto \cos{\frac{\theta}{2}} u_3, \qquad
    r\delta \theta \propto \sin{\frac{\theta}{2}} u_3.
\end{equation}
Thus, linear active phonon modes with superexchange require a mode with non-zero projection on the $u_3$ direction. There is an exception for collinear ($\theta=0$) superexchange paths, as shown in Fig. \ref{fig:Fig4_superexchange}(c), where $u_2$ and $u_3$ are equivalent and lead to linear spin-phonon coupling where the phase of the displacement is irrelevant.

\begin{figure}[!htb]
\includegraphics[width=1.0\columnwidth]{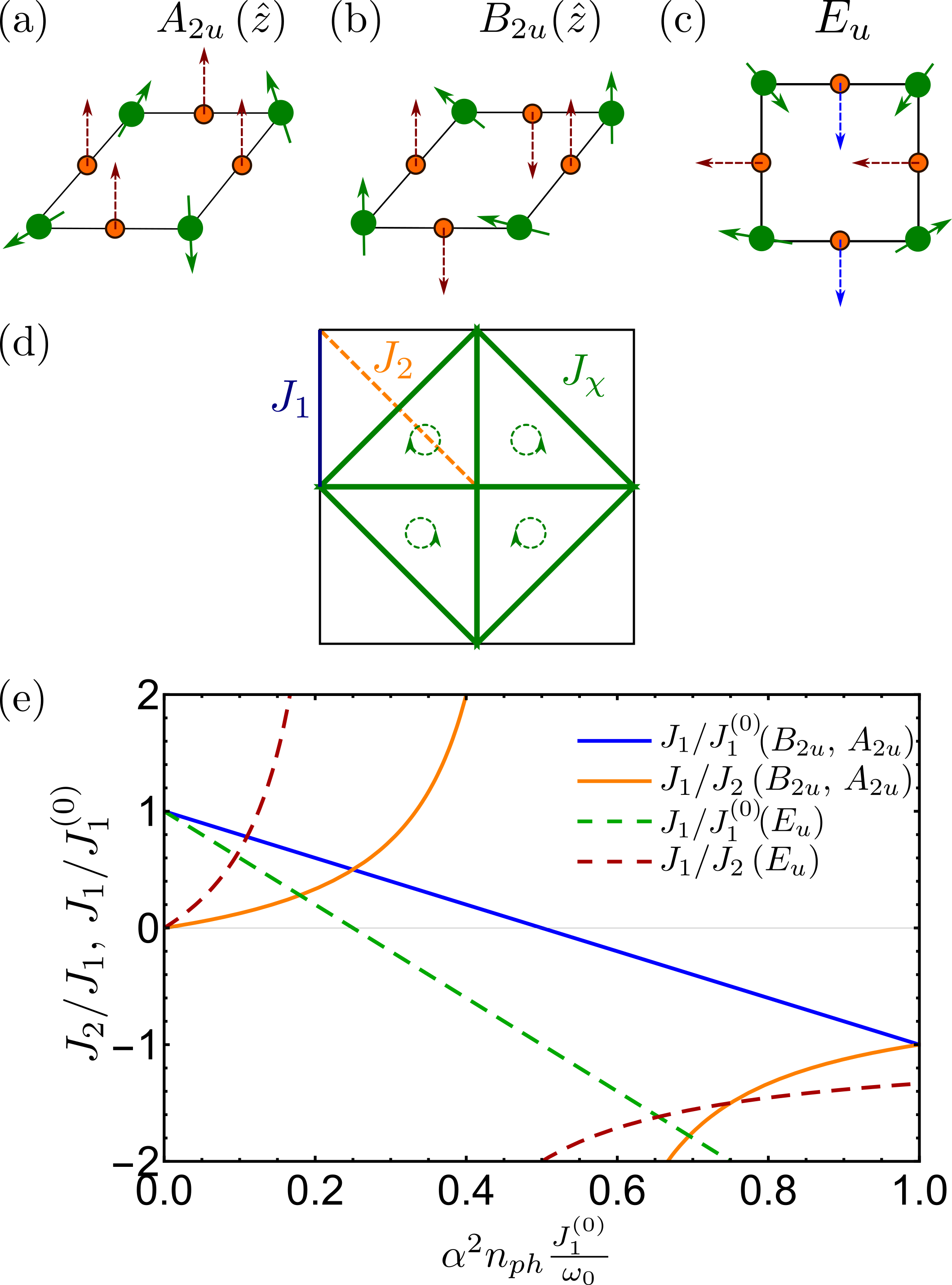}
\caption{Edge-stuffed square lattice phonons and phonon pumped superexchange couplings. Three optical phonon modes have linear spin-phonon coupling: (a) $A_{2u}$, (b) $B_{2u}$ and (c) $E_u$. (d) The phonon pumping of these modes generates nearest and next-nearest neighbor interactions from initial $S=1/2$ nearest neighbor couplings. The polarization of the $E_u$ mode in the collinear case does not generate chiral couplings, but the non-collinear variation would allow polarization to generate a $C_4$ staggered chirality. (e) Nearest and next-nearest neighbor couplings as a function of $\alpha^2 n_{ph}J_1^{(0)}/\omega_0$. $A_{2u}$ and $B_{2u}$ are indistinguishable in the collinear case, while pumping $E_u$ gives a larger prefactor. 
\label{fig:Fig5_square_couplings}}
\vspace{-0.cm}
\end{figure}

Now we apply these general spin-phonon coupling rules to the edge-stuffed square lattice with linear superexchange paths. The unit cell has three sites: one metal and two anions. Three optical modes are spin-phonon active, as shown in Fig. \ref{fig:Fig5_square_couplings}(a)-(c): two singly degenerate, $A_{2u}$ and $B_{2u}$, and a doubly degenerate mode, $E_u$. In the frame of stationary metal ions, the eigenvectors can be fully described by the anion displacements, 
\begin{equation}
\label{eq:sqmodes}
   \!\hat{\epsilon}_{A_{2u}}\!\!=\! \frac{1}{\sqrt{2}}\!\begin{pmatrix}
0 \\
\hat{z} \\
\hat{z} \\
\end{pmatrix}\!\!, \;  \hat{\epsilon}_{B_{2u}}\!\!=\! \frac{1}{\sqrt{2}}\!\begin{pmatrix}
0 \\
\hat{z} \\
-\hat{z} \\
\end{pmatrix}\!\!, \;
\hat{\epsilon}_{E_{g}}^{R/L}\!\!=\! \frac{1}{\sqrt{2}}\!\begin{pmatrix}
0 \\
\hat{x} \\
\pm i\hat{y} \\
\end{pmatrix}\!
\end{equation}
in the metal-ligand-ligand ion basis, where $\hat{z}$ is perpendicular to the plane of the lattice.

We now calculate the effective spin couplings. We expect to potentially generate nearest- and next-nearest exchange couplings, chiral terms, and four-spin terms that might include ring exchange. Explicit calculation finds that the four spin terms vanish, while $J_1$, $J_2$ and a staggered chiral term $J_{\chi}$ are allowed, as shown in Fig. \ref{fig:Fig5_square_couplings}(d). With linear superexchange paths, the phase of the eigenmodes is irrelevant, so even polarized $E_u$ pumping does not generate chiral terms, similar to what was found for Floquet pumping the square lattice \cite{Quito2021b}. To obtain $J_{\chi}$, the ions must be statically shifted away from the linear path. The phase insensitivity also means the effects of $A_{2u}$ and $B_{2u}$ are indistinguishable. The phonon driven couplings are:

\begin{align}\label{eq:sqcouplings}
    J_1^{\left(A_{2u}/B_{2u}\right)}&=J_1^{(0)}\left(1-2\alpha^2 n_{ph}\frac{J_1^{(0)}}{\omega_0}\right),\cr
    J_2^{\left(A_{2u}/B_{2u}\right)}&=2J_1^{(0)} \alpha^2 n_{ph}\frac{J_1^{(0)}}{\omega_0},\cr
J_1^{\left(E_u\right)}&=J_1^{(0)}\left(1-4\alpha^2 n_{ph}\frac{J_1^{(0)}}{\omega_0}\right),\cr
    J_2^{\left(E_u\right)}&=4J_1^{(0)} \alpha^2 n_{ph}\frac{J_1^{(0)}}{\omega_0},
\end{align}
as shown in in Fig. \ref{fig:Fig5_square_couplings}(e).  Many features are similar to the 
honeycomb case. As before, it is possible to drive $J_2/J_1$ into interesting regimes with realistically accessible coupling strengths.  The square lattice has a potential spin liquid region for $J_2/J_1 \gtrsim .4$ \cite{Jiang2012, Liu2018, Becca2020}.  An additional feature of the square lattice is the possibility to select the type of phonon mode. In this case, the difference between $E_u$ and $A_{2u}/B_{2u}$ is purely quantitative, with $E_u$ giving twice the effect because there are twice as many modes.

\subsection{Anisotropic models on the honeycomb lattice\label{sec:anisotropic}}

Finally, we consider spin-phonon pumping on several anisotropic spin models on the honeycomb lattice, chosen because of its simple optical phonon modes, and because it hosts the exactly solvable Kitaev honeycomb spin liquid \cite{Kitaev2006}. We consider models of the form,
\begin{equation}\label{eq:genSz}
    H_s=
    \sum_{i l, jl'}J_{ij,l l'} S_i^{l}S_j^{l'},
\end{equation}
where $l = x,y,z$ denotes spin components, allowing $J_{ij,l l'}$ to be anisotropic and/or bond-dependent. We begin with the XYZ model, with $J_{1,x} \neq J_{1,y}\neq J_{1,z}$ and show how the anisotropy changes under phonon pumping; then we treat the Kitaev honeycomb model and show how the anisotropy can be enhanced to drive the system from a gapless to a gapped spin liquid; finally we consider the experimentally relevant Kitaev-Heisenberg model \cite{Jackeli2009, Chaloupka2013, Witczak2014, Gohlke2017} and show how spin-phonon pumping can selectively enhance Kitaev terms over Heisenberg ones.

\begin{figure}[!htb]
\includegraphics[width=0.95\columnwidth]{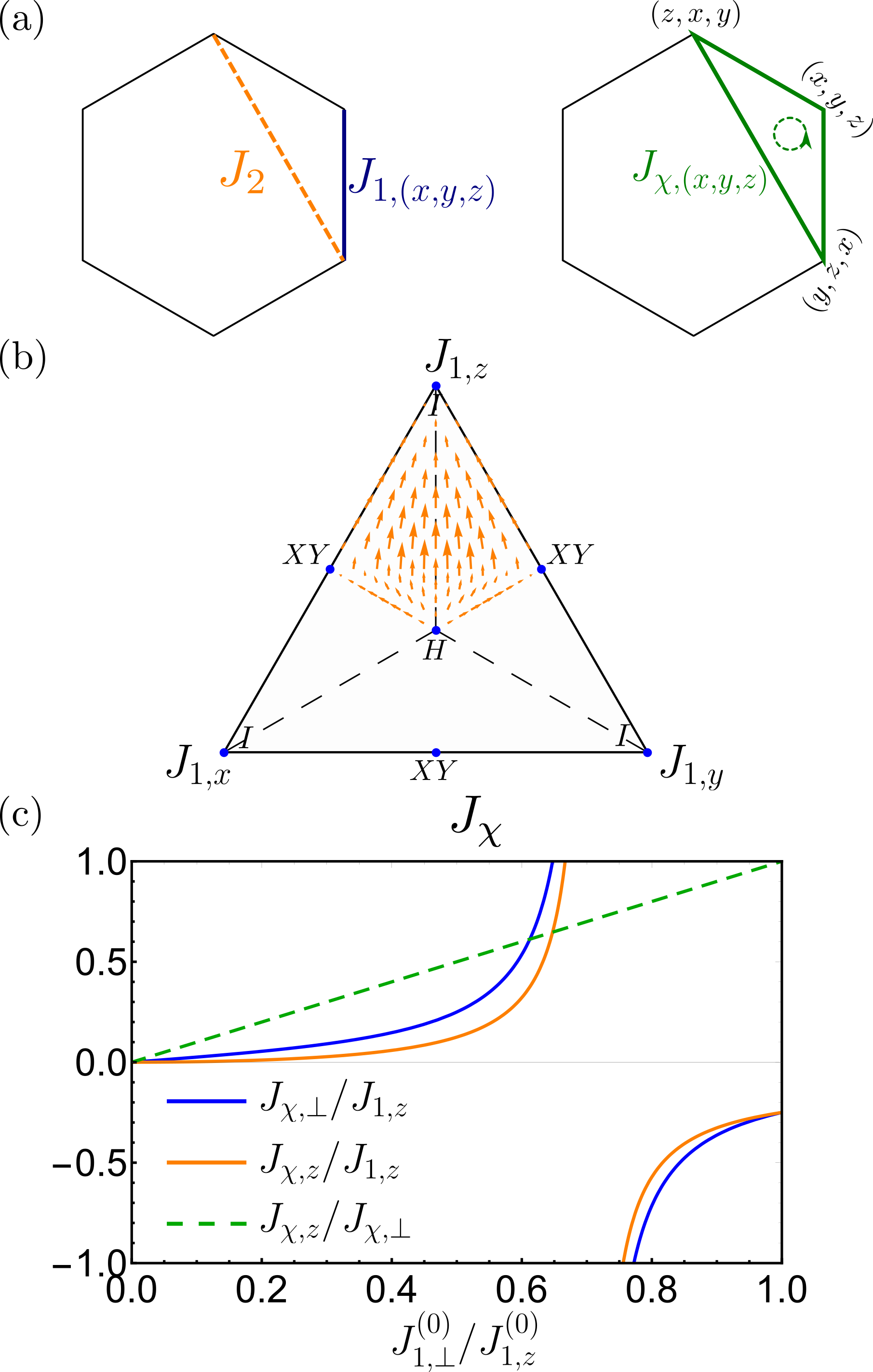}
\caption{Phonon pumping of the XYZ honeycomb lattice. (a) Phonon pumping of the $S=1/2$ XYZ model generates anisotropic $J_1$ and isotropic $J_2$ couplings for the unpolarized case (left) and additional anisotropic chiral couplings in the phonon polarized case (right). (b) Phonon pumped anisotropic couplings in barycentric coordinates. Each point corresponds to the original (unpumped) couplings, while the arrow gives the direction and magnitude of the coupling change under phonon pumping, with couplings changing along straight lines. Ising points (vertices) are stable invariant points against phonon pumping; the Heisenberg point (center) is an unstable invariant point; and XY points (midpoints of sides) are marginally stable invariant points. (c) Anisotropic pumped chiral couplings as a function of initial anisotropy for the XXZ case, with $\omega_0/J_{1, z}^{(0)}=\beta=1/2$ and $\alpha^2 n_{ph}=1/3$. \label{fig:Fig8_anisHeis_res}}
\vspace{-0.cm}
\end{figure}

\subsubsection{XYZ model\label{sec:anisHeis}}

We consider the XYZ model, 
\begin{equation}\label{eq:XYZ}
    H_s=\sum_{\langle i j\rangle}
    \sum_{l=x,y,z}J_{1,l}^{(0)}S_i^{l}S_j^{l},
\end{equation}
and spin-phonon coupling with the same XYZ anisotropy,
\begin{equation}\label{eq:HspXYZ}
H_{sp}=\sum_{\langle ij\rangle}\sum_{l=x,y,z} J_{1,l}^{(0)} \lambda_{l} u_{ij} S_i^{l}S_j^{l},
\end{equation}
a direct extension of Eq. (\ref{eq:Hsp}). For simplicity, we take $\lambda_{l}=\lambda$, making the original exchange anisotropy and spin-phonon anisotropy identical. Besides simplifying the analysis, we expect this to hold for direct exchange. The anisotropic $J$'s motivate the definition of $\alpha$ as,
\begin{equation}\label{eq:alphaZ}
    \alpha^2=\frac{\lambda^2}{2M_i \omega_{0} }\frac{J_{1,z}^{(0)}}{\omega_{0}},
\end{equation}
where we define $J_{1,z}^{(0)}$ to be the largest initial coupling.

We proceed with the analysis along the lines of Sec. \ref{sec:LFtransf} on the honeycomb lattice with $E_{2u}$ optical phonon described in Sec. \ref{sec:HCheis}, and extract the effective exchange couplings from the phonon distribution dependent terms. The nested commutator structure introduces the same constraints as the isotropic case. In particular, as shown in Fig. \ref{fig:Fig8_anisHeis_res}(a), the generated two-spin terms include first and second nearest neighbor exchange; three-spin terms generate chiral couplings; and the four-spin terms cancel. The generated interactions are:
\begin{align}\label{eq:XYZcouplings}
    J_{1,l}&=J_{1,l}^{(0)}\left[1-3\alpha^2 n_{ph}\frac{J_{1,z}^{(0)}}{\omega_0}\frac{\sum_{m, m\neq l}\left(J_{1,k}^{(0)}\right)^2}{\left(J_{1,z}^{(0)}\right)^2}\right],\cr
    J_2& =3J_{1,z}^{(0)}\alpha^2 n_{ph}\frac{J_{1,z}^{(0)}}{\omega_0} \frac{J_{1,x}^{(0)}J_{1,y}^{(0)}J_{1,z}^{(0)}}{\left(J_{1,z}^{(0)}\right)^3}, \cr
     J_{\chi,l}&=3\sqrt{3}J_{1,z}^{(0)} \alpha^2 n_{ph}\beta|\epsilon_{lmn}|\frac{J_{1,m}^{(0)} J_{1,n}^{(0)}}{\left(J_{1,z}^{(0)}\right)^2}
\end{align}
where $J_{1,l}$ are obtained by cyclic permutations of the numerator, $l,m = x,y,z$ and the extra $J_{1,z}^{(0)}$s are due to defining $\alpha$ in terms of $J_{1,z}^{(0)}$.

The nearest-neighbor and chiral terms are generically anisotropic, while the generated $J_2$ is forced to be isotropic; is only generated when all three couplings are non-zero; and is largest for isotropic couplings, as shown  in Fig. \ref{fig:Fig8_anisHeis_res}(c).  In general, phonon pumping \emph{enhances} the existing anisotropy.  We show this in the anisotropy diagram in Fig. \ref{fig:Fig8_anisHeis_res}(b), where each point in barycentric coordinates specifies the original (unpumped) coupling constant ratios. The arrow at each point denotes the direction and strength of the coupling change under spin phonon pumping. There are three types of ``stable'' anisotropy points corresponding to the Heisenberg, XY and Ising models, which are the only cases where phonon pumping does not increase the anisotropy.  

Lastly, polarized pumping generates anisotropic chiral couplings, with the following Hamiltonian:
\begin{align}\label{eq:chiraltermsXYZ}
\delta H_{\chi}=&\sum_{\substack{\langle ijk \rangle\\lmn = x,y,z}}\epsilon_{lmn} J_{\chi, l} S_{i}^{l} S_{j}^{m}S_{n}^{z}
\end{align}
with $\langle ijk \rangle$ denoting the triangle in Fig. \ref{fig:Fig8_anisHeis_res}(a).
The ratio of different chiral components is directly related to the original  anisotropy, as shown in Fig. \ref{fig:Fig8_anisHeis_res}(c) for the XXZ case ($J_{\chi, x}=J_{\chi, y}=J_{\chi, \perp}$). 

We expect many of these features to hold for general anisotropic models, including the types of couplings generated, their dependence on the phonon pumping, and the increase in anisotropy with pumping.  This increase arises from the structure of pumped terms coming from the nested commutators, where, for example, the $y$ component nearest-neighbor term comes from commuting $x$ and $z$ terms. This structure means the smallest $J_1$ couplings are suppressed the most, as they have the largest prefactors in phonon-pumped terms. Further neighbor couplings are also generically expected to be suppressed with higher anisotropy, as they require different components participating on different bonds in the commutator. More broadly, the spin-phonon pumping terms vanish for effectively classical models, which have no non-vanishing commutators, as is seen in the Ising limit here. Quantum fluctuations will generically aid spin-phonon pumping.

\subsubsection{Kitaev honeycomb model\label{sec:Kitaev}}

\begin{figure}[!htb]
\includegraphics[width=1.0\columnwidth]{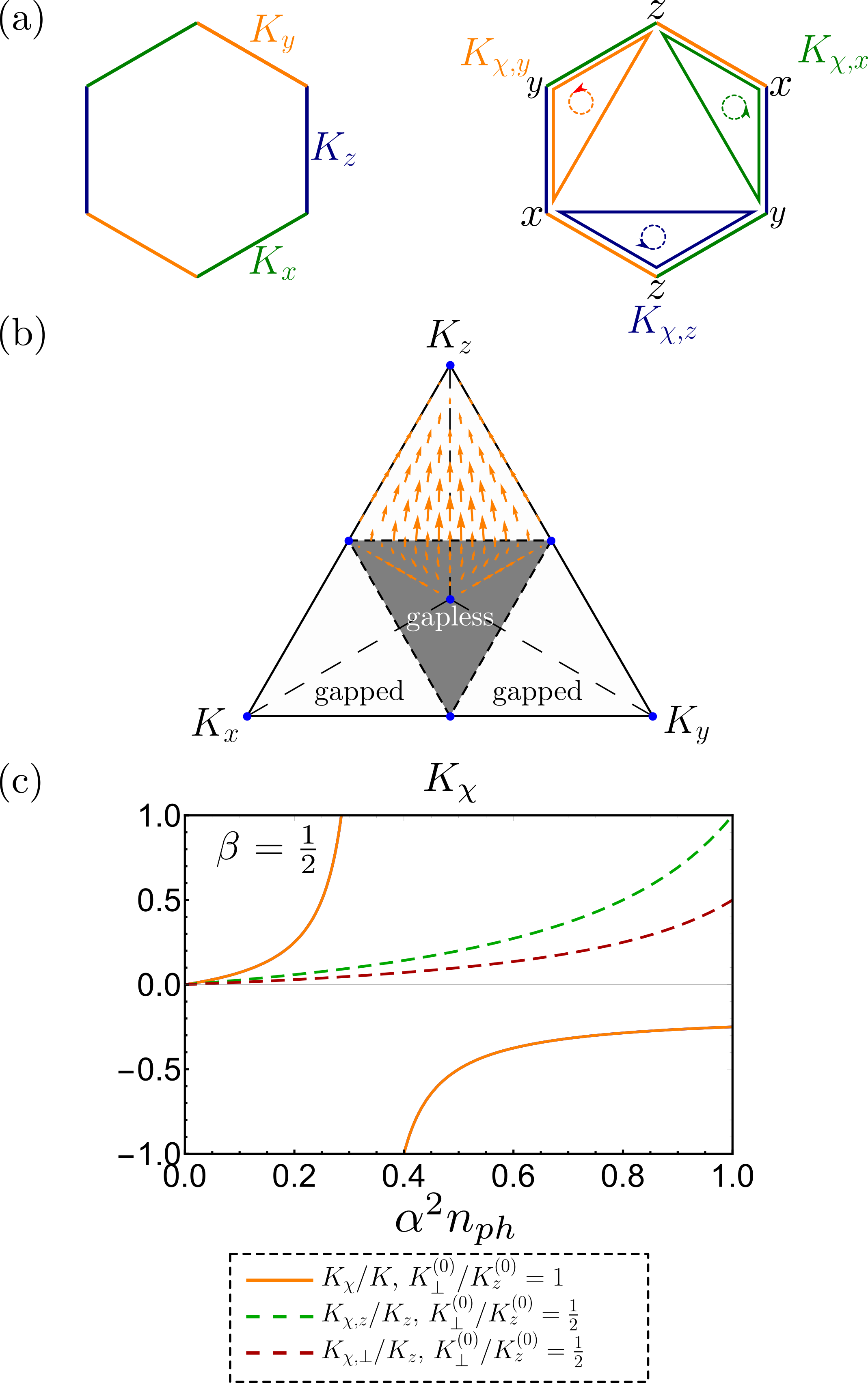}
\caption{Phonon pumping of the Kitaev model. (a) Kitaev couplings pumped by unpolarized phonons lead to no further neighbour couplings generated (left), while polarized phonons give rise to chiral Kitaev couplings (right). (b)  Phonon pumped Kitaev couplings in barycentric coordinates, with the same notation as Fig. \ref{fig:Fig8_anisHeis_res}. The coupling change is qualitatively the same as for the $XYZ$ model. Phonon pumping could allow tuning between gapless and gapped Kitaev spin liquids. (c) Chiral couplings as a function of $\alpha^2n_{ph}$ at $\omega_0/K^{(z,0)}=\beta=1/2$ for several initial anisotropies. \label{fig:Fig9_Kitaev_res}}
\vspace{-0.cm}
\end{figure}

Next, we consider the Kitaev honeycomb model \cite{Kitaev2006}, characterized by bond dependent anisotropic interactions on the three nonequivalent nearest-neighbor honeycomb bonds (labeled by $x,y,z$), $K_{x,y,z}$:
\begin{equation}\label{eq:KitaevHs}
    H_s=
    \sum_{l=x,y,z}\sum_{\langle ij\rangle_l} K_l^{(0)}S_{i,l}S_{j,l},
\end{equation}
as shown in Fig. \ref{fig:Fig9_Kitaev_res}(a). This model is exactly solvable \cite{Kitaev2006}, with a ground state that is a gapless Kitaev spin liquid for $K_z<K_x+K_y$ (and cyclic permutations), and a gapped $Z_2$ liquid otherwise. This Kitaev exchange can be realized, along with additional exchange interactions, in strongly spin-orbit coupled edge-sharing octahedral materials like $\alpha-$RuCl$_3$ \cite{Banerjee2016} and several iridate families \cite{Jackeli2009, Okamoto2007, Modic2014}.  The presence of additional Heisenberg or $\Gamma$ exchange terms not only spoil the exactly solvable nature, but generically destabilize the spin liquid phases for sufficiently large magnitudes, making selective pumping of different interactions highly desirable.

We first focus on the pure anisotropic Kitaev honeycomb model and consider the $E_{2u}$ optical mode. The generated interactions [see Fig. \ref{fig:Fig9_Kitaev_res}(a)] include nearest-neighbor Kitaev exchanges and bond dependent chiral couplings with polarized pumping. The nested commutator structure cannot generate second neighbor interactions by phonon pumping a pure Kitaev Hamiltonian.  The bond dependent chiral coupling takes the form:
\begin{align}\label{eq:chiraltermsKitaev}
\delta H_{\chi}=&\sum_{\substack{\langle ijk \rangle'\\lmn = x,y,z}}\epsilon_{lmn}K_{\chi,l} S_{i,l} S_{j,m}S_{k,n},
\end{align}
where $\langle ijk \rangle'$ denotes the triangle with a central $i$ vertex between $m$ and $n$ Kitaev bonds. The same chiral coupling arises in the Kitaev model under other time-reversal breaking perturbations, like magnetic field \cite{Kitaev2006}. The expressions for effective couplings are reminiscent of corresponding XYZ model expressions,
\begin{align}\label{eq:Kitaev_couplings}
    K_x=&K_x^{(0)}\left[1-\frac{3}{2}\alpha^2 n_{ph}\frac{K_z^{(0)}}{\omega_0}\frac{\left(K_y^{(0)}\right)^2+\left(K_z^{(0)}\right)^2}{\left(K_z^{(0)}\right)^2}\right],\nonumber \\
     K_{\chi,x}=&3\sqrt{3} K_x^{(0)} \alpha^2 n_{ph}\beta \frac{K_y^{(0)} K_z^{(0)}}{\left(K_z^{(0)}\right)^2},
\end{align}
with $\alpha^2\equiv(\lambda^2/2M_i \omega_{0}) K_z^{(0)}/\omega_{0}$; the $y,z$ expressions are cyclic permutations of the above. The anisotropy in nearest-neighbor couplings is again enhanced by spin-phonon pumping and looks nearly identical to the XYZ case, as shown in Fig. \ref{fig:Fig9_Kitaev_res}. Therefore, spin-phonon pumping can potentially change the character of the spin-liquid state from gapless to gapped. The chiral coupling anisotropy mirrors the original anisotropy, and the size of the chiral coupling decreases with increased initial anisotropy, as shown in Fig. \ref{fig:Fig9_Kitaev_res}(c).

\subsubsection{Kitaev-Heisenberg model\label{sec:KitaevHeis}}

\begin{figure*}[!htb]
\includegraphics[width=0.8\textwidth]{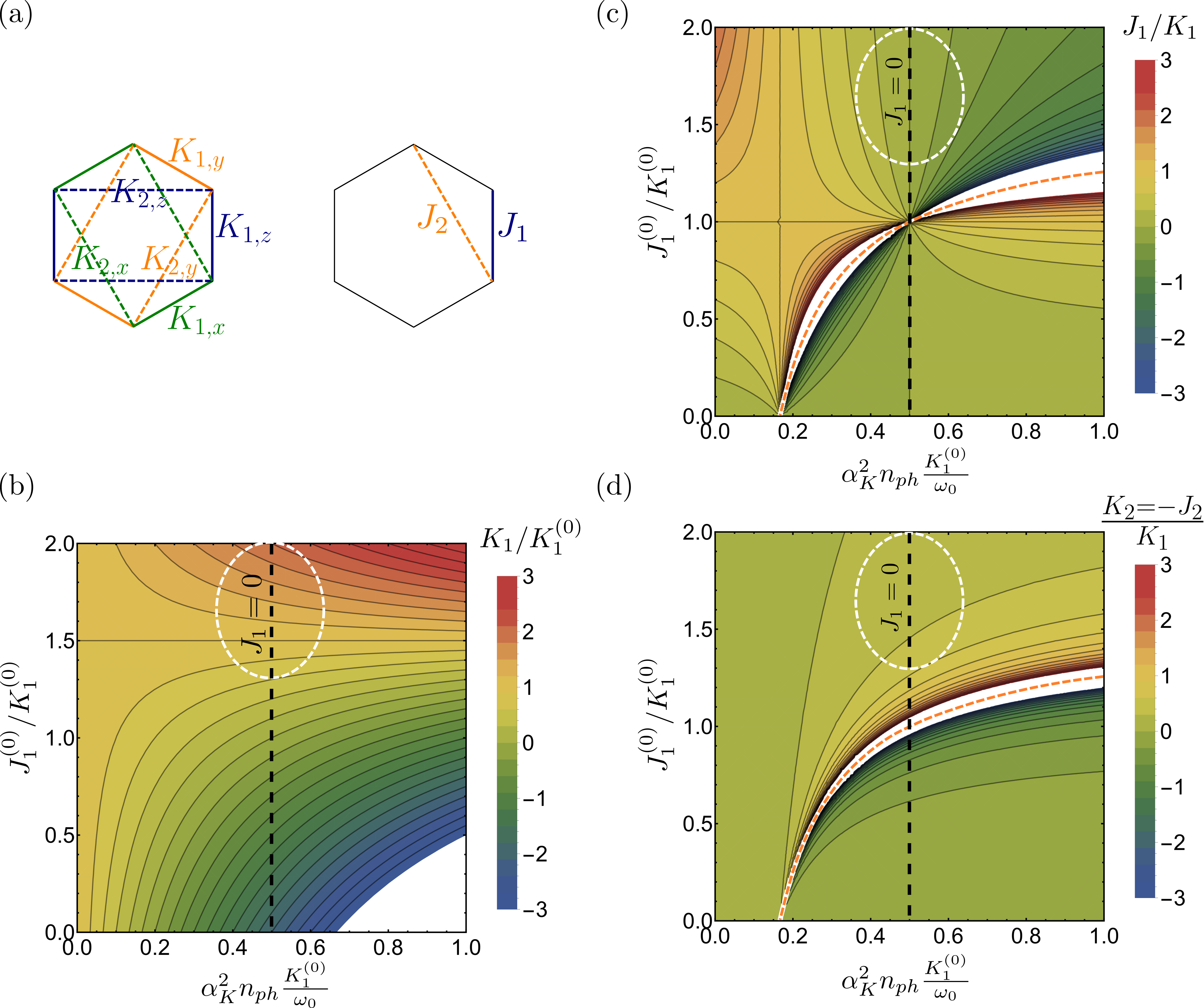}
\caption{Phonon pumping of Kitaev-Heisenberg model. (a) Both first and second neighbor Kitaev and Heisenberg couplings are generated by unpolarized phonon pumping of Kitaev-coupled phonon modes. Pumped nearest neighbor Kitaev coupling (b), nearest neighbor Heisenberg coupling (c) and next nearest neighbor Heisenberg/Kitaev couplings as a function of $\alpha_K^2 n_{ph}K_1^{(0)}/\omega_0$ and the initial Heisenberg-Kitaev coupling ratio, $J_1^{(0)}/K_1^{(0)}$. The $J_1 = 0$ (dashed, black) line is shown on all three plots, with $K_1 = 0$ (dashed, red) present in (c)-(d). Regions of interest with large $|K_1|$ and small $J_1$ and $J_2, K_2$ are circled (dashed, white).}
\label{fig:Fig10_KitaevHeis_res}
\vspace{-0.cm}
\end{figure*}

Finally, we consider an isotropic Kitaev-Heisenberg model \cite{Jackeli2009, Chaloupka2013, Gohlke2017},
\begin{equation}\label{eq:KHham}
    H_s=J_1^{(0)} \sum_{\langle i j\rangle}\vect{S}_{i}\cdot \vect{S}_{j} + K_1^{(0)}\sum_{l=x,y,z}\sum_{\langle ij\rangle_l} S_{i,l}S_{j,l},
\end{equation}

The general spin-phonon Hamiltonian includes independent Heisenberg and Kitaev spin-phonon couplings, 
\begin{equation}\label{eq:HspKH}
H_{sp}=\!\!\sum_{\langle ij\rangle}J_{1}^{(0)} \lambda_{H} u_{H,ij} \vect{S}_{i}\cdot \vect{S}_{j} +\!\!\!\!\sum_{\substack{l=x,y,z\\\langle ij\rangle_la}}\!\!\! K_1^{(0)} \lambda_{K} u_{K,ij} S_{i,l}S_{j,l}.
\end{equation}
Here, $\lambda_{K(H)}$ and $u_{K(H),ij}$ are corresponding spin-phonon couplings and relevant phonon displacements for the Kitaev and Heisenberg interactions, respectively, as defined in Eg. (\ref{eq:uij}). In the real materials, Heisenberg and Kitaev exchange arise from different physical processes: Heisenberg primarily arises from direct exchange involving metal-metal hopping, while Kitaev comes from anion mediated superexchange \cite{Chaloupka2013}. Thus,  the spin-phonon couplings and relevant phonon modes will be significantly different.  For simplicity, we set $\lambda_{H}=0$; this could, for example, represent a phonon mode that primarily displaces the octahedral ligand anions, thus only affecting the Kitaev-generating superexchange.  Results are qualitatively similar for the other limit, $\lambda_{K}=0$, but these are more favorable for enhancing $K_1/J_1$.

We now perform the spin-phonon analysis for the $E_{2u}$ phonon, considering only the unpolarized case.  This generates only two-spin interactions, with changed nearest neighbor couplings ($K_1$ and $J_1$) and emergent second neighbor couplings ($K_2$ and $J_2$), as shown in Fig. \ref{fig:Fig10_KitaevHeis_res}. The second neighbor Kitaev couplings are bond dependent, with the $x$ component on the second neighbor sites connected by $y$ and $z$ bonds (and cyclically permuted). The expressions are similar to previous examples, but now $K_1$ additionally depends on $J_1^{(0)}/K_1^{(0)}$,
\begin{align}\label{eq:KHcouplings}
    K_1=&K_1^{(0)}\left[1-2\alpha_K^2 n_{ph}\frac{K_1^{(0)}}{\omega_0}\left(3-2\frac{J_1^{(0)}}{K_1^{(0)}} \right)\right],\cr
     J_1=&J_1^{(0)}\left[1-2\alpha_K^2 n_{ph}\frac{K_1^{(0)}}{\omega_0}\right],\cr
     K_2=&-J_2=\frac{1}{2}\alpha_K^2 n_{ph}\frac{K_1^{(0)}}{\omega_0} J_1^{(0)},
\end{align}
with $\alpha_K^2=(\lambda_K^2/2M_i \omega_{0}) K^{(0)}_1/\omega_{0}$. 

The relevant coupling constant ratios are shown as a function of both tuning parameters: phonon occupancy and the initial $J_1^{(0)}/K_1^{(0)}$ ratio in Fig. \ref{fig:Fig10_KitaevHeis_res}(b)-(d). The 2D parameter space contains an interesting regime in which spin-phonon pumping selectively enhances $K_1$ Kitaev interactions, while suppressing all other couplings. Tuning a material into this regime would induce a Kitaev spin liquid. In this case, the regime arises at intermediate pumped spin-phonon coupling with either large or small initial $J_1^{(0)}/K_1^{(0)}$. The existence of a region of comparatively enhanced $K_1$ is enabled by the nonlinearity of the spin-phonon pumped magnetic interactions in Eq. (\ref{eq:KHcouplings}), which arises from the interplay between the non-commuting Kitaev and Heisenberg terms in the original spin Hamiltonian, Eq. (\ref{eq:KHham}). In real materials, the phonons and exchange interactions will be more complicated, but we expect there will still be specific phonon modes that can be pumped to selectively enhance the Kitaev interactions.

\section{Conclusions \label{sec:conclusions}}

We have developed a general treatment of the leading quantum effects of dynamically generated lattice fluctuations on magnetic interactions, and have shown that non-equilibrium spin-phonon pumping can be used to generate a rich range of magnetic interactions, including additional exchange couplings, chiral fields, and enhanced magnetic anisotropies. The result is a generic, easy to interpret set of rules describing which types of magnetic interactions are accessible by phonon pumping, based on the geometry of the lattice, magnetic exchange mechanism and geometry of the pumped phonon modes. This approach has the potential to drive existing materials into very interesting model regimes with experimentally accessible parameters. These possibilities include accessing spin liquid regimes on the triangular, honeycomb and square Heisenberg models; tuning the Kitaev honeycomb anisotropy between gapless and gapped spin liquids; and reducing the problematic Heisenberg interactions in the Kitaev-Heisenberg model.  The non-equilibrium IR phonon occupations required should be accessible with current experimental capabilities, while the relevant frequencies are generically in the sub-gap regime for magnetic insulators, meaning that electronic heating should be minimal, with phonon heating similarly expected to be weak \cite{Kennes2017}.

Our approach is perturbative and expected to work well in the diabatic regime where the phonons are fast compared to the spins. In the future, full quantum calculations on toy models would be helpful to evaluate the validity of perturbation theory, while direct dynamical simulations can evaluate the diagonal phonon approximation. These studies are particularly important for extending this analysis to the adiabatic regime \cite{Ferrari2021}. 

Finally, to apply this approach in real materials, quantitative predictions require knowing the starting magnetic interactions and associated spin-phonon coupling strengths, either from modeling or ab initio calculations. Testing these predictions not only requires being able to pump the system into the desired state, but the ability to measure the state on the relevant experimental time scales, which will typically require optical techniques to follow phase transitions, magnetic excitations or spin liquid signatures \cite{Potter2013,Pilon13,Pustogow2018,Colbert14}.  Non-linear optical spectroscopy \cite{Zhang2014, Nicoletti2016, Gianetti2016} can potentially both implement phonon pumps and distinguish the emergent spin states through coherent 2D spectroscopy of excitations \cite{Lu2018, Wan2019, Choi2020, Qiang2023}.

\vspace{12pt}
\begin{acknowledgments}
We acknowledge stimulating discussions with Ana-Marija Nedi\' c,  Yihua Qiang and Victor L. Quito. This work was supported by NSF Grant No. DMR-1555163.
\end{acknowledgments}

\appendix
\section{Kagom\'{e} lattice Heisenberg model\label{sec:kagome}}

Here, we consider the mono-atomic kagom\'{e} lattice, which hosts multiple spin-phonon active optical modes, allowing for mode-dependent phonon pumping of a diverse set of exchange and chiral interactions, in contrast to the Honeycomb lattice example from the main text \ref{sec:HCheis}.

\begin{figure}[!htb]
\includegraphics[width=1.0\columnwidth]{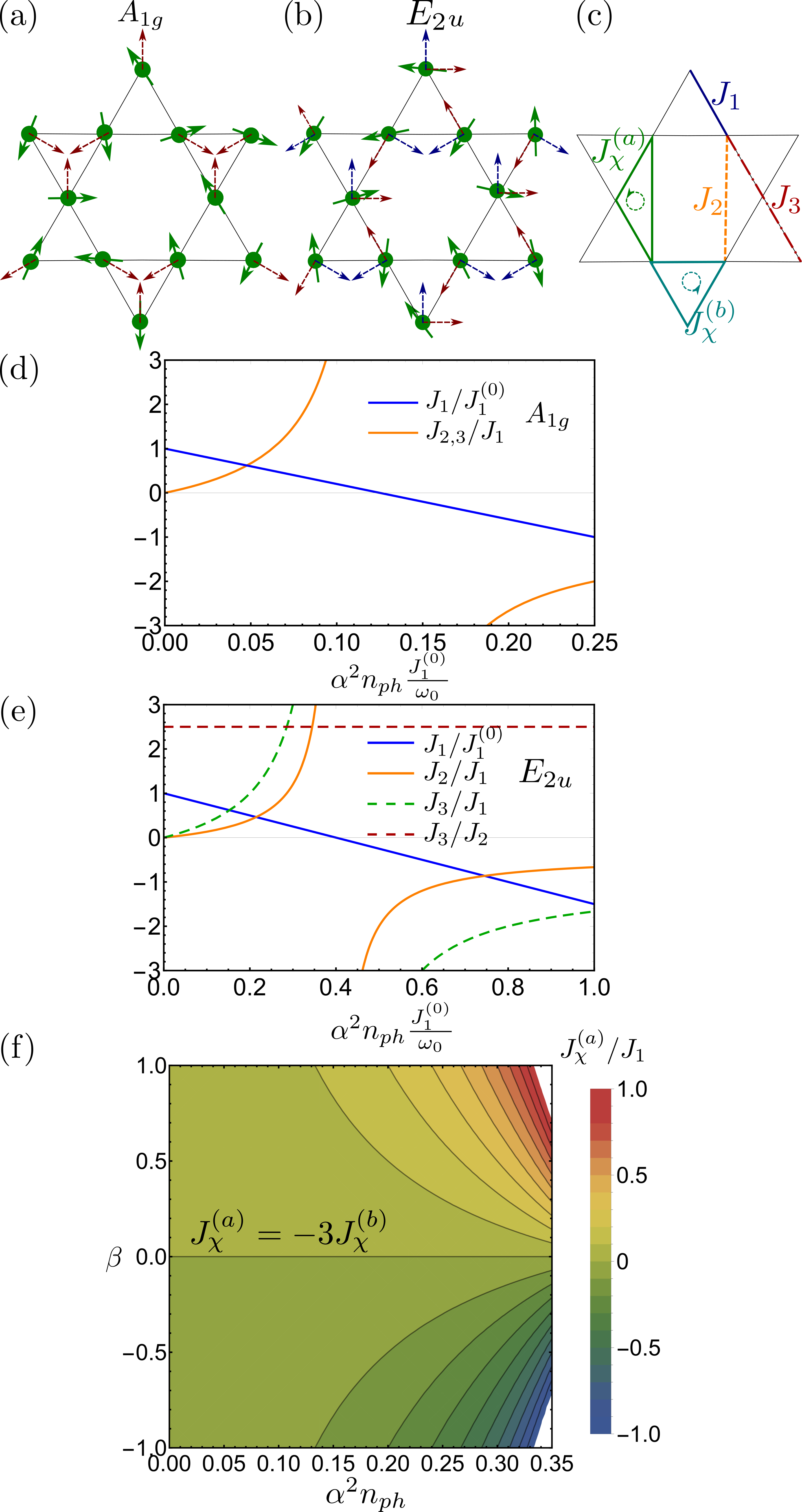}
\caption{Kagom\'{e} lattice phonons and modified exchange couplings. The linearly spin-coupled kagom\'{e} phonons are the (a) $A_{1g}$ and (b) $E_{2u}$ modes. (c) Phonon pumping of the $S=1/2$ nearest neighbor model generates $J_1$, $J_2$ and $J_3$ couplings, and two additional chiral terms for the polarized $E_{2u}$ case. Exchange couplings as a function of the relevant spin-phonon coupling are shown for (d) the $A_{1g}$ mode and (e) the $E_{2u}$ mode. Divergences are simply due to $J_1$ vanishing. (f) Chiral couplings for polarized $E_{2u}$ pumping with $\omega_0/J_1^{(0)}=.5$. \label{fig:Fig7_Kagome_res}}
\vspace{-0.cm}
\end{figure}

The spin-phonon active modes on kagom\'{e} lattice are those with non-zero projections onto the bond directions. There are three such modes: the singly degenerate $A_{1g}$ and the doubly degenerate $E_{2u}$, shown in Fig. \ref{fig:Fig7_Kagome_res}(a)-(b). In the three-atom basis, the modes are,
\begin{align}\label{eq:Kgmodes}
   \hat{\epsilon}_{A_{1g}}&= \frac{1}{\sqrt{3}}\begin{pmatrix}
\hat{y} \\
\frac{\sqrt{3}}{2}\hat{x}-\frac{1}{2}\hat{y} \\
-\frac{\sqrt{3}}{2}\hat{x}-\frac{1}{2}\hat{y} \\
\end{pmatrix}, \cr 
\hat{\epsilon}_{E_{2u}}^{R/L}&= \frac{1}{\sqrt{6}}\begin{pmatrix}
\hat{x} \pm i\hat{y} \\
\xi(\hat{x} \pm i\hat{y})\\
\xi^{*}(\hat{x} \pm i\hat{y}), \\
\end{pmatrix}
\end{align}
where $\xi=e^{2\pi i/3}$.

Applying the spin-phonon pumping formalism for these modes, we obtain exchange and chiral interactions, as shown in Fig. \ref{fig:Fig7_Kagome_res}. The nested commutator rules now allow two distinct further neighbor exchanges, $J_2$ and $J_3$. Again, the four-spin terms are not forbidden, but are explicitly zero. The expressions for the relevant pumped couplings take the following forms:
\begin{align}\label{eq:Kgcouplings}
    J_1^{\left(A_{1g}\right)}=&J_1^{(0)}\left(1-8\alpha^2 n_{ph}\frac{J_1^{(0)}}{\omega_0}\right),\cr
    J_2^{\left(A_{1g}\right)}=&J_3^{\left(A_{1g}\right)}=8 J_1^{(0)}\alpha^2 n_{ph}\frac{J_1^{(0)}}{\omega_0},\cr
    J_1^{\left(E_{2u}\right)}=&J_1^{(0)}\left(1-\frac{5}{2}\alpha^2 n_{ph}\frac{J_1^{(0)}}{\omega_0}\right),\cr
    J_2^{\left(E_{2u}\right)}=& J_1^{(0)} \alpha^2 n_{ph}\frac{J_1^{(0)}}{\omega_0},\cr
    J_3^{\left(E_{2u}\right)}=&\frac{5}{2} J_1^{(0)} \alpha^2 n_{ph}\frac{J_1^{(0)}}{\omega_0},\cr
    J_{\chi}^{(E_{2u},a)}=&-3J_{\chi}^{(E_{2u},b)}=J_1^{(0)} \alpha^2 n_{ph}\beta
\end{align}
Here and in what follows, $\alpha^2$ and $n_{ph}$ denote the coupling constant and pumped phonon occupancies for each of the respective modes ($A_{1g}$ and $E_{2u}$). There are two distinct chiral terms, $J_\chi^{(a)}$ and $J_\chi^{(b)}$ [see Fig. \ref{fig:Fig7_Kagome_res}(c)] for the $E_{2u}$ mode, which have a fixed ratio, $J_\chi^{(a)} = -3 J_\chi^{(b)}$; note that the negative sign here means that the chiralities add constructively in the overall unit cell.  The modified exchange and chiral interactions are shown as a function of the relevant drive parameters in Fig. \ref{fig:Fig7_Kagome_res}(d)-(f), where again the nearest-neighbor $J_1$ is reduced, while $J_2$ and $J_3$ are enhanced.  Notably, the $A_{1u}$ mode forces $J_2 = J_3$, while the $E_{2u}$ mode forces $J_3 = 5/2 J_2$; a mixed drive might obtain intermediate $J_3/J_2$ ratios.  

The nearest-neighbor kagom\'{e} lattice is intrinsically highly frustrated, with a likely $Z_2$ spin liquid ground state \cite{Jiang2008, Yan2011}. The different paths in the $J_1-J_2-J_3$ kagom\'{e} phase diagram achievable with spin phonon pumping provide for the possibility of selectively tuning between a $Z_2$ spin liquid around $J_2=J_3=0$, and a chiral spin liquid \cite{Gong2015} or alternative frustrated classical phases. The range of $J_3/J_2$ where the chiral liquid is stabilized is $\sim 0.75-4$, with $J_2/J_1\gtrsim 0.05$, well within the range of accessibility of our simple spin-phonon pumping model.

\section{Superexchange on the triangular lattice\label{sec:Triangular}}

\begin{figure}[!htb]
\includegraphics[width=.95\columnwidth]{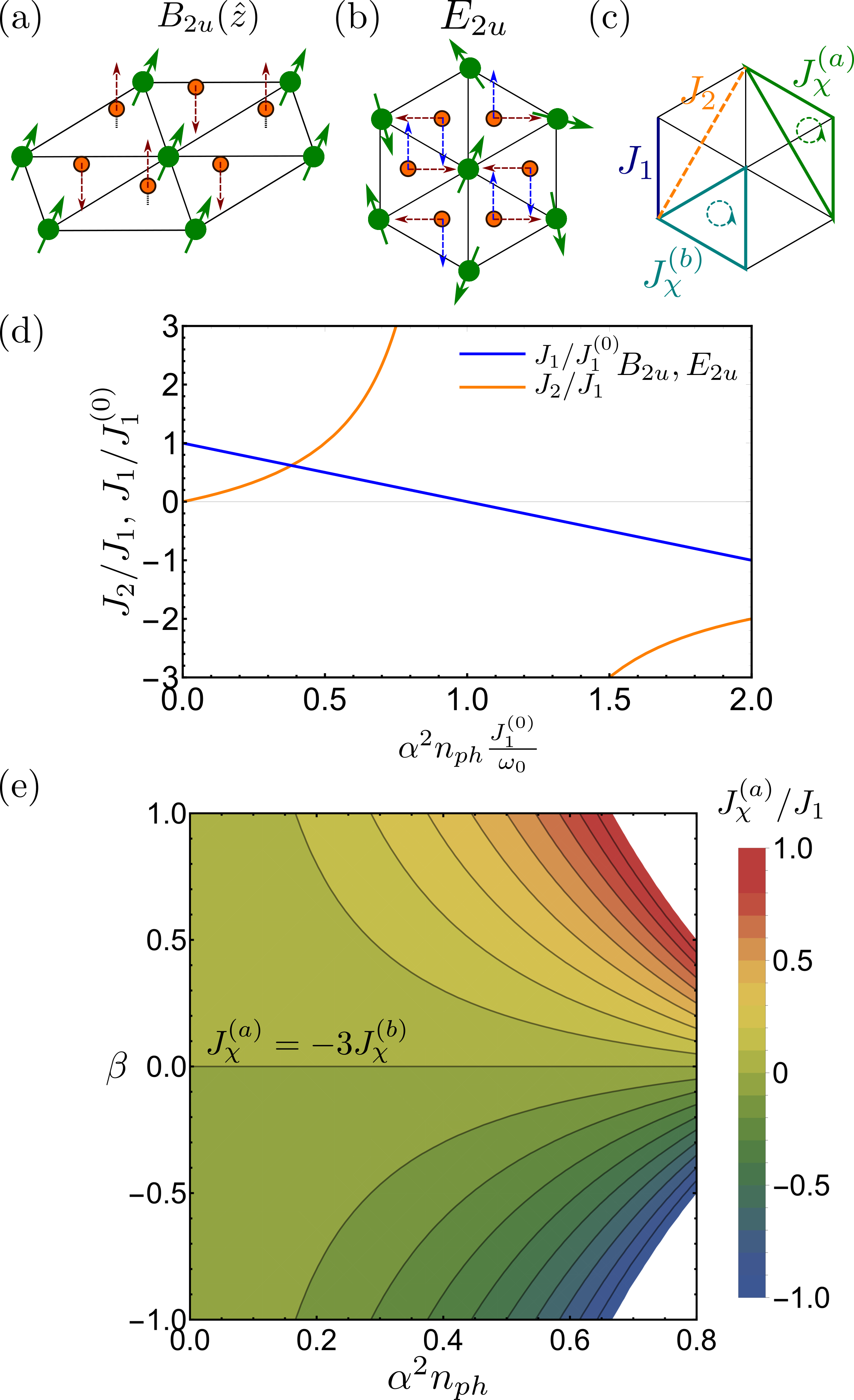}
\caption{Stuffed triangular lattice phonons and pumped exchange couplings. Linearly coupled phonon modes on the stuffed triangular lattice must be staggered between the triangles and are thus limited to (a) $B_{2u}$ and (b) $E_{2u}$. (c) Phonon pumping the $S=1/2$ nearest neighbor Heisenberg model modifies the $J_1$ and $J_2$ couplings in the unpolarized case and adds two distinct chiral terms in the polarized case. (d) Exchange couplings as a function of relevant coupling, which are equivalent for the two modes. (e) Chiral couplings for polarized $E_{2u}$ pumping with $\omega_0/J_1^{(0)}=1/2$.  $J_{\chi}^{(a)}=-3J_{\chi}^{(b)}$ irrespective of pumping or polarization. \label{fig:Fig6_triangular_res}}
\vspace{-0.cm}
\end{figure}

We supplement the discussion of Sec.~\ref{sec:SEpaths} by exploring a triangular lattice stuffed with superexchange mediating ligand anions at the centers of the triangles, shown in Fig. \ref{fig:Fig6_triangular_res}(a). In this case, the non-collinear superexchange paths lead to chiral interactions for polarized $E_{2u}$ phonon pumping.

This stuffed triangular lattice has three sites per unit cell, one metal and two anions. Each nearest-neighbor interaction is the sum of two independent superexchange paths, potentially allowing additional cancellation between the two paths. As a result, only two modes are spin-phonon active, $B_{2u}$ and $E_{2u}$. The eigenmode displacements shown in Fig. \ref{fig:Fig6_triangular_res}(a)-(b) are represented in stationary frame of the metal atoms by,
\begin{align}\label{eq:trmodes}
   \hat{\epsilon}_{B_{2u}}&= \frac{1}{\sqrt{2}}\begin{pmatrix}
0 \\
\hat{z} \\
-\hat{z} \\
\end{pmatrix}, \qquad   
\hat{\epsilon}_{E_{2u}}^{R/L}&= \frac{1}{\sqrt{2}}\begin{pmatrix}
0 \\
\hat{x} \\
\pm i\hat{y} \\
\end{pmatrix}.
\end{align}

The possible spin-phonon generated interactions are shown in Fig. \ref{fig:Fig6_triangular_res}(c). The  two spin interactions include changes in the nearest- and the generation of next-nearest-neighbor exchange. Polarized phonon pumping can generate two distinct chiral terms, while the four-spin terms, including ring exchange again exactly vanish. The resulting magnetic couplings are,
\begin{align}\label{eq:trcouplings}
    J_1^{\left(B_{2u}/E_u\right)}&=J_1^{(0)}\left(1-\alpha^2 n_{ph}\frac{J_1^{(0)}}{\omega_0}\right),\cr
    J_2^{\left(B_{2u}/E_u\right)}&=J_1^{(0)}\alpha^2 n_{ph}\frac{J_1^{(0)}}{\omega_0},\cr
    J_{\chi}^{\left(E_u,a\right)} &=-3J_{\chi}^{\left(E_u,b\right)}=J_1^{(0)}\alpha^2 n_{ph}\beta.
\end{align}
As before, $J_1$ is reduced.  Here, both modes have equivalent effects on $J_1$ and $J_2$, although only the polarized $E_{2u}$ pumping induces chiral terms, $J_{\chi}^{(a)}=-3J_{\chi}^{(b)}$. The relative sign means the chiralities add constructively.  On the triangular lattice\cite{Wietek2017}, the chiral terms add to a single effective chirality $\left|J_{\chi}^{(eff)}\right|=\left|J_{\chi}^{(a)}\right|+\left|J_{\chi}^{(b)}\right|$ for any three or four-sublattice magnetic orders and some spin liquids.  This relation does not hold more generally, but the qualitative effect should still be additive.

The generated $J_2/J_1$ and chiral couplings are shown in Fig. \ref{fig:Fig6_triangular_res}(d)-(e). The frustrated triangular lattice means that $J_2/J_1\gtrsim 0.1$ can lead to either a $Z_2$ or Dirac quantum spin liquid \cite{Zhu2015, Gong2019}, which could be readily accessed with spin-phonon pumping. A distinct chiral spin liquid was found for $J_{\chi}^{(b)} \gtrsim 0.03$ \cite{Wietek2017}. As the two chiral interactions generated by spin-phonon pumping are additive, we expect that the chiral liquid could be stabilized with accessible polarized $E_u$ pumping.
\newpage


%


\end{document}